\documentclass{aa}

\journalname{}

\usepackage{graphicx}
\usepackage{txfonts}
\usepackage{booktabs}
\usepackage{hyperref}
\usepackage{amsmath}
\usepackage{mathrsfs}
\usepackage{amssymb}
\usepackage{mathtools}

\newcommand{\norm}[1]{\lVert#1\rVert}

\bibpunct{(}{)}{;}{a}{}{,}

\begin{document}

\title{Capture and escape from the 2:1 resonance between Ariel and Umbriel in a fast-migration scenario of the Uranian system}
\titlerunning{Capture and escape from the 2:1 Ariel-Umbriel resonance in a fast-migration scenario}

\author{M. Rossi
          \inst{1}
          \and
          G. Lari\inst{1}
          \and
          M. Falletta\inst{1}
          \and
          C. Grassi\inst{1}
          \and
          S.R.A. Gomes\inst{2}
          \and
          M. Saillenfest\inst{3}
          }

\institute{Department of Mathematics, University of Pisa, Largo Bruno Pontecorvo 5, 56127 Pisa, Italy\\
\email{mattia.rossi@dm.unipi.it}\and 
Anton Pannekoek Institute for Astronomy, University of Amsterdam, Science Park 904, 1098, XH, Amsterdam, The Netherlands \and
LTE, Observatoire de Paris, Université PSL, Sorbonne Université, Université de Lille, LNE, CNRS, 61 Avenue de l’Observatoire, 75014 Paris, France}

\date{Received --; accepted --}

  \abstract 
   {Recent measurements and theoretical developments suggest that Uranus may exhibit a higher tidal dissipation than previously assumed. This enhanced dissipation leads to a faster orbital migration of its five major moons: Miranda, Ariel, Umbriel, Titania, and Oberon. Consequently, when studying their orbital evolution, resonant encounters that have always been discarded in previous works need to be included. In particular, Ariel’s fast migration implies that the crossing of the 2:1 mean motion resonance with Umbriel is extremely likely and it could have occurred in recent times (within the last one billion years). Capture into this strong resonance would have induced significant tidal heating within Ariel, possibly explaining its resurfacing. Therefore, in this work, we aim to explore the orbital history of the Uranian moons in a context of fast tidal migration, including the crossing of the 2:1 mean motion resonance between Ariel and Umbriel. For small initial eccentricities, we confirm that the moons are always captured into this resonance, which can potentially persist indefinitely. As the system is not currently involved in any mean motion resonance, we investigated possible dynamical mechanisms for exiting the 2:1 resonance. We show that the resonance could have been broken by a further resonant encounter with Titania. We analyzed the crossing of the 4:2:1 and 3:2:1 resonant chains, and looked for the parameter space where the probability of escaping the resonance is maximized. In particular, taking a dissipative parameter of Ariel $k_{2,2}/Q_2<10^{-3}$, the passage through the 3:2:1 resonance succeeds in disrupting the 2:1 mean motion resonance between Ariel and Umbriel in more than $60\%$ of our numerical experiments. While the crossing of the three-body resonant chain produces a general excitation of the orbital elements of all moons, a fraction of our simulations results in final low eccentricities and inclinations, which can eventually match the current orbital features of the system. As the proposed orbital history requires specific ranges of the dissipative parameters for the system, it will be possible to validate (or disprove) this scenario with the data of the next space mission to Uranus.}

   \keywords{celestial mechanics,  planets and satellites: dynamical evolution and stability}

   \maketitle
   
\section{Introduction}
\label{sec:intro}

The planet Uranus hosts five regular satellites (Miranda, Ariel, Umbriel, Titania, and Oberon) whose dynamical history is far from being fully determined. Unlike Jovian and Saturnian regular moons, these satellites are not involved in any mean motion resonance, although it is plausible that during their past evolution they crossed different resonant configurations (e.g., \citealp{DERMOTT-etal_1988}). Some of them also show wide signs of resurfacing episodes, which are probably evidence of past resonant interactions between the moons \citep{PEALE_1999}. In addition, some Uranian moons may harbor (or have harbored) subsurface oceans under their icy crust \citep{CASTILLO-etal_2023}, so there is great interest in reconstructing the thermal-orbital history of these satellites. We do not know how the regular satellites of Uranus have been formed in the first place, and whether or not their formation process is related to the large tilt of Uranus's spin axis \citep{MORBIDELLI-etal_2012,RUFU-CANUP_2022,SALMON-CANUP_2022}. Understanding how they may have obtained their current orbits would provide key information on these matters as well.

In the Planetary Science and Astrobiology Decadal Survey 2023–2032, a mission to Uranus was identified as the next priority Flagship project. The proposed Uranus Orbiter and Probe \citep{SIMON-etal_2023} is currently under development and will arrive at the system of Uranus not earlier than 2040. The mission will collect a great variety of data on the planet and its moons, thanks to instruments and experiments onboard the spacecraft. As the last probe to visit the Uranus system was Voyager 2 in 1986, this new space mission offers a unique opportunity to acquire precise in-situ observations, essential to improve our knowledge on the current state and the history of Uranus and its satellites (see, e.g., \citealp{PETRICCA-etal_2025}).

Although the current orbits and masses of the moons have already been accurately determined \citep{JACOBSON_2014}, the orbital evolution of the Uranian moons is still unclear. The current orbital configuration of the system is the result of a billion-year evolution which depends on many factors. Tidal dissipation within the planet, whose magnitude is set by the inverse of the quality factor $Q_U$, induces orbital migration of the satellites, producing the crossing of orbital resonances. Types and epochs of resonant encounters depend on the value of $Q_U$ and the initial orbits of the moons, which are not known. In a series of papers, Tittemore and Wisdom performed an extensive analysis of mean motion resonances that could have played a role in shaping the Uranian moon system \citep{TITTEMORE-WISDOM_1988,TITTEMORE-WISDOM_1989,TITTEMORE-WISDOM_1990}. Assuming a low tidal dissipation scenario for Uranus (i.e., high $Q_U$), as inferred by classical tidal theories of giant planets \citep{GOLDREICH-SOTER_1966}, they explored the main resonances expected to be encountered by the satellites. They obtained that the passage through the 3:1 mean motion resonance between Miranda and Umbriel could be responsible for the current high inclination of Miranda, and that the passage through the 5:3 mean motion resonance between Ariel and Umbriel could explain the relatively high free eccentricities of the satellites. Moreover, they found that if Ariel and Umbriel had crossed the 2:1 resonance, these moons would have been almost surely captured, with no possibility of disrupting the resonance. Therefore, in order to avoid this resonant encounter, the authors suggested an upper limit on the orbital migration of the satellites by setting the value of $Q_U$ larger than $11\,000$.

The passage and temporary capture into mean motion resonances allow one to explain not only the current relatively high eccentricities and inclinations of some moons, but also part of their observed geological features. The surfaces of Miranda and Ariel, and also Titania, show signs of past resurfacing episodes, which are evidence of past high internal heating. Since radiogenic heating alone cannot be responsible for such catastrophic events, tidal heating has been invoked as the primary source of past resurfacing. As tidal dissipation within satellites depends on the moons' orbital eccentricities (e.g., \citealp{PEALE-etal_1979}), for generating enough tidal heating it is necessary that some mean motion resonances were active in the past and forced the moons' eccentricities to values higher than those observed today (see, e.g., \citealp{DERMOTT-etal_1988,TITTEMORE_1990,TITTEMORE-WISDOM_1988}).

In this context, recent works revisited the crossing of the 5:3 resonance between Ariel and Umbriel, being the closest to the current orbital configuration and the most likely candidate for the latest orbital excitation of the system. Considering a quality factor of Uranus of the order of $10\,000$ (i.e., near the lower bound set by previous works), this crossing should have happened between half and a few billion years ago. Through numerical simulations, \citet{CUK-etal_2020} found that temporary capture into this resonance would have triggered a chaotic phase that excited the eccentricities and inclinations of all moons in the system, including those not directly involved in the mean motion resonance. This process could explain why the free eccentricities of Umbriel and Ariel are significantly different from zero, as well as the formation of young
coronae on Miranda's surface, which are evidence of recent tectonic activities. Even more strikingly, during this chaotic phase, Miranda’s inclination increased from nearly zero to about 4 degrees, providing an alternative mechanism to account for its unusual value. However, the inclinations of the other satellites also increased significantly, in contrast to their currently low values. To avoid this issue, \citet{GOMES-CORREIA_2024b} investigated conditions to cross the 5:3 resonance without being captured. They found that avoiding temporary capture into resonance is possible, but it requires that the eccentricity of Ariel before the resonant encounter should have been larger than about $0.015$, probably as the result of a previous unidentified resonant interaction between the satellites.

All these previous works assumed traditional estimates for the tidal dissipation within Uranus inherited from the foundational work of \citet{GOLDREICH-SOTER_1966}. However, recent theoretical studies and measurements suggest that tidal dissipation within giant planets could be much higher than previously thought \citep{FULLER-etal_2016,LAINEY-etal_2020}, resulting in a sustained high migration rate of their moons, or at least of some of them. For Saturn, a quality factor as low as $100$ has been measured, with consequences on the entire dynamical evolution of its system \citep{CUK-etal_2024}. In the case of Uranus, \citet{NIMMO_2023} recently proposed a value of the quality factor of Uranus of about $1\,000$ to explain the resurfacing of moons through resonant captures (see also \citealp{MEYER-WISDOM_2007}). Specifically, the author showed that if Ariel had been locked into a low-order mean motion resonance with Umbriel and $Q_U$ had been around $1\,000$ at that epoch, the tidal heating within Ariel would have reached hundreds of GW, matching the prediction of the geophysical estimates obtained from the analysis of Ariel's surface features \citep{PETERSON-etal_2015}. Furthermore, \citet{JACOBSON-PARK_2025} presented the first measurement of this parameter, obtaining $Q_U=678\pm 231$. This value needs to be taken with caution, as it was obtained assuming the same quality factor of Uranus for all moons, while we know from observations of the Saturn system that $Q$ of the planet can be very different at the various orbital frequencies of the satellites \citep{LAINEY-etal_2017,LAINEY-etal_2020}. Nevertheless, it is a clear indication of strong tidal dissipation within Uranus.

A low value of Uranus's quality factor strongly affects the dynamical history of the satellites. The first implication is that the crossing of the 5:3 resonance between Ariel and Umbriel is very recent, i.e., less than 50 million years ago. More importantly, a fast migration of Ariel makes the 2:1 resonance between Ariel and Umbriel almost impossible to avoid. While all previous orbital histories of the Uranian moons relied on avoiding this resonance (see \citealp{TITTEMORE-WISDOM_1990}), its inclusion may drastically change our understanding of the orbital and thermal history of the Uranian moons. Therefore, in light of a high tidal dissipation scenario for Uranus, we aim to study the evolution of the Uranus system as it evolves through this 2:1 resonance and to investigate whether such a crossing could be compatible with its current orbital configuration. Possibly being the strongest resonance encountered by the entire system, its effects could allow us to explain many orbital and geological features observed in the Uranian moons. Nevertheless, it is necessary to prove that the resonance can eventually be broken (despite the conclusions of the early work by \citealp{TITTEMORE-WISDOM_1990}) in order to reach the current non-resonant configuration of the system.

The paper is structured as follows: in Sect.~\ref{sec:dynmo}, we present the orbital dynamical model and the setup we use for our investigation; in Sect.~\ref{sec:2bres}, we study the capture of Ariel and Umbriel into the 2:1 mean motion resonance and its effects on their orbits; in Sect.~\ref{sec:break}, we show a dynamical mechanism to break down the 2:1 resonance involving a three-body resonant encounter with Titania. Finally, in Sect.~\ref{sec:discu} and Sect.~\ref{sec:concl}, we discuss and summarize the results we obtained. Further details and technical aspects of the model presented in Sect.~\ref{sec:dynmo} are given in App.~\ref{sec:appA}.

\begin{table*}
    \centering
    \caption{Physical parameters and mean orbital elements of the Uranian moon system used in the numerical simulations. Masses and radii are from \citet{JACOBSON_2014} and references therein contained; normalized moments of inertia are taken from \citet{GOMES-CORREIA_2024a}, where for Uranus $\zeta_U=0.225$; tidal Love numbers and quality factors of the satellites, as well as constant (i.e., no resonance-locking) quality factors of the planet, are assigned suitable values for the purpose of the present parametric study (see Sects.~\ref{subsec:tides} and \ref{subsec:resloc}), with $k_{2,U}=0.3$ for Uranus \citep{STIXRUDE-etal_2021}; Ariel is always assumed in resonance locking according to Eq.~\eqref{eq:Qreslock}, whereas Titania is or is not depending on the type of resonant encounter (two-body or three-body, see Sects.~\ref{sec:2bres} and \ref{sec:break}). Semi-major axes, eccentricities and inclinations are from \citet{GOMES-CORREIA_2024a}, whereas longitudes of pericenter and ascending node are obtained by averaging the JPL ephemerides at J2000 epoch \citep{JACOBSON_2014}.}
    \begin{tabular}{c c c c c c}
    \toprule\toprule
     & Miranda & Ariel & Umbriel & Titania & Oberon \\
    \midrule
    $m_i$ [$m_U$] & $7.4552178\times10^{-7}$ & $1.4405272\times10^{-5}$ & $1.4686589\times10^{-5}$ & $3.9169073\times 10^{-5}$ & $3.5437548\times 10^{-5}$ \\
    $R_i$ [$R_U$] & $9.3900\times10^{-3}$ & $2.2736\times10^{-2}$ & $2.2876\times10^{-2}$ & $3.0866\times 10^{-2}$ & $2.9790\times 10^{-2}$\\
    $\zeta_i$ & 0.327 & 0.320 & 0.342 & 0.326 & 0.310\\
    $k_{2,i}$ & 0.03 & 0.15 & 0.03 & 0.03 & 0.03\\
    $Q_i$ & $10\,000$ & $\in[50,400]$ & $1\,000$ & $1\,000$ & $1\,000$\\
    $Q_{U,i}$ & $10\,000$ & Eq.~\eqref{eq:Qreslock} & $10\,000$ & $10\,000$ or Eq.~\eqref{eq:Qreslock} & $10\,000$\\
    $a_i$ [$R_U$] & 5.080715 & 7.470167 & 10.406589 & 17.069604 & 22.827536\\
    $e_i$ & 0.00135 & 0.00122 & 0.00394 & 0.00123 & 0.00140\\
    $I_i$ [deg] & 4.4072 & 0.0167 & 0.0796 & 0.1129 & 0.1478\\
    $\varpi_i$ [deg] & 108.8614 & 253.4145 & 184.7195 & 49.7581 & 95.0536\\
    $\Omega_i$ [deg] & 304.5699 & 113.0462 & 39.3105 & 190.7227 & 244.4884\\
    \bottomrule
    \end{tabular}
    \label{tab:param}
\end{table*}

\section{Dynamical model}
\label{sec:dynmo}

Averaged dynamical models (with respect to fast angles) are a standard technique in celestial mechanics and astrodynamics for investigating the long-term evolution of satellite systems (\citealp{MURRAY-DERMOTT_2000}). This approach is primarily motivated by computational efficiency, as the dynamics must be analyzed over timescales of hundreds of millions or even billions of years, making it reasonable to remove fast periodic contributions and only look at the underlying long-term dynamics. In our case, we want to perform a statistical and parametric study of the capture into, and subsequent exit from, the 2:1 resonance between Ariel and Umbriel. Since such an analysis requires running hundreds of simulations, repeatedly using full $N$-body integrations would be computationally prohibitive. Moreover, fast periodic terms would possibly mask the long-term signals we are looking for. Analyzing the output of direct $N$-body simulations would therefore require the use of filtering techniques, which are not necessary if we directly work with an appropriately averaged model.

\subsection{Resonant averaged Hamiltonian}
\label{subsec:avgham}

In the following, we consider a five-satellite model in an equatorial Uranus-centric reference frame, formulated within a Hamiltonian setting as a perturbation of five independent Kepler problems. Therefore, the Hamiltonian function is the sum of the Keplerian term $\mathcal{H}_K$ and some perturbative terms. We consider two generic two-body mean motion resonances: Ariel and Umbriel close to a $(p_1+q_1)/p_1$ commensurability, and Umbriel and Titania close to a $(p_2+q_2)/p_2$ commensurability, where $p_1,p_2,q_1,q_2$ are nonzero integers such that the pairs $(p_1+q_1)$, $p_1$ and $(p_2+q_2)$, $q_2$ are co-prime. The resulting third commensurability between Ariel and Titania is established from the two previous relations as $(p_1+q_1)(p_2+q_2)/(p_1p_2)$. This leads to a three-body resonant chain between Ariel, Umbriel and Titania. In our model, we include the perturbation due to Uranus's oblateness and the direct and indirect gravitational interactions among the moons. We checked that the inclusion of the Sun's attraction (direct and indirect) does not introduce significant changes in the dynamics (see also \citealp{GOMES-CORREIA_2024a}).

We start from the Hamiltonian expressed in Cartesian coordinates and then we expand it in elliptical elements up to second order in the eccentricities $e_i$ and inclinations $I_i$ of the moons (App.~\ref{subsec:expham}), where the index $i=1,\ldots,5$ indicates the $i$-th moon in order of increasing distance from the planet. Then, the Hamiltonian is averaged over the mean longitudes $\lambda_i$, except for the resonant combinations $(p_1+q_1)\lambda_3-p_2\lambda_2$, $(p_2+q_2)\lambda_4-p_2\lambda_3$, and $((p_1+q_1)(p_2+q_2)/g)\lambda_4-(p_1p_2/g)\lambda_2$, where $g=\mathrm{GCD}((p_1+q_1)(p_2+q_2),p_1p_2)$ and GCD denotes the greatest common divisor. For the purpose of the present work, we will consider two cases (see Sect.~\ref{sec:break}): the `Laplace resonance' 4:2:1, corresponding to $p_1=p_2=q_1=q_2=1$ (2:1 Ariel-Umbriel, 2:1 Umbriel-Titania), and the Laplace-like resonance 3:2:1, corresponding to $p_1=q_1=q_2=1$, $p_2=2$ (2:1 Ariel-Umbriel, 3:2 Umbriel-Titania). 

The final form of the Hamiltonian reads
\begin{equation}
    \label{eqn:ham}
    \mathcal{H}=\mathcal{H}_K+\mathcal{H}_O+\mathcal{H}_I+\mathcal{H}_D\;,
\end{equation}
where the Keplerian term is
\begin{equation}
    \label{eqn:hamkep}
    \mathcal{H}_K=-\sum_{i=1}^5\frac{\mathcal{G} m_U m_i}{2a_i}\;,
\end{equation}
the oblateness term is
\begin{equation}
    \label{eqn:hamobl}
    \mathcal{H}_O=\sum_{i=1}^5\frac{\mathcal{G}m_Um_i}{a_i}J_2\left(\frac{R_U}{a_i}\right)^2\frac14\left(12s_i^2-3e_i^2-2\right)\;,
\end{equation}
the indirect term of the mutual gravitational perturbation is
\begin{equation}
    \label{eqn:hamI}
    \mathcal{H}_I=\begin{cases}
        \begin{aligned}
        &\frac{\beta_3a_3n_3}{m_U}\bigg(\beta_2a_2n_2e_3\cos(2\lambda_3-\lambda_2-\varpi_3)\\
        &\qquad\; +\beta_4a_4n_4e_4\cos(2\lambda_4-\lambda_3-\varpi_4)\bigg)
        \end{aligned}& \text{(case 4:2:1)}\\
        \begin{aligned}
        &\frac{\beta_2a_2n_2}{m_U}\bigg(\beta_3a_3n_3e_3\cos(2\lambda_3-\lambda_2-\varpi_3)\\
        &\qquad\; +9\beta_4a_4n_4e_4^2\cos(3\lambda_4-\lambda_2-2\varpi_4)\bigg)
        \end{aligned}& \text{(case 3:2:1)}
    \end{cases}\;,
\end{equation}
while the direct term is
\begin{equation}
    \label{eqn:hamD}
    \mathcal{H}_D=\mathcal{H}_{D,\text{sec}}+\mathcal{H}_{D,\text{2:1},23}+\begin{cases}
        \mathcal{H}_{D,\text{2:1},34} & \text{(case 4:2:1)} \\
        \mathcal{H}_{D,\text{3:2},34}+\mathcal{H}_{D,\text{3:1},24} & \text{(case 3:2:1)}
    \end{cases}\;.
\end{equation}
In the above expressions, $\mathcal{G}$ is the gravitational constant; $R_U$ and $m_U$ are the radius and mass of Uranus, respectively;  $s_i=\sin(I_i/2)$ are functions of the inclinations $I_i$; $m_i$, $a_i$, $\varpi_i$, $\mu_i=\mathcal{G}(m_U+m_i)$, $\beta_i=m_Um_i/(m_U+m_i)$, and $n_i=\sqrt{\mu_i/a_i^3}$ are the mass, semi-major axis, longitude of pericenter, gravitational parameter, reduced mass, and mean motion of the $i$-th satellite, respectively. Finally, $J_2$ is the second zonal harmonic coefficient of Uranus, whose value is set to $3.510685\times10^{-3} $\citep{JACOBSON_2014}.

Eq.~\eqref{eqn:hamD} contains both secular and resonant terms for each three-body chain configuration. The specific terms depend on the type of resonance and moons involved, which are indicated by the labels. Their detailed expressions are reported in App.~\ref{subsec:HD}. Note that in the 4:2:1 resonance case, we do not include resonant contributions involving the couple Ariel-Titania, as the corresponding terms appear only at third order in eccentricities and inclinations. 

\subsection{Canonical variables}
\label{subsec:canvb}

In order to derive the equations of motion from the Hamiltonian~\eqref{eqn:ham}, we choose a convenient set of canonical variables. As is customary, we adopt the Poincaré resonant variables, which are well-defined for $e_i=s_i=0$. To this end, we begin with the modified Delaunay canonical elements:
\begin{equation}
\label{eqn:modDel}
    \begin{cases}
		\Lambda_i=\beta_i\sqrt{\mu_i a_i}\\
		\Gamma_i=\Lambda_i\left(1-\sqrt{1-e_i^2}\right)\\
		Z_i=\Lambda_i\sqrt{1-e_i^2}\left(1-\cos I_i\right)
	\end{cases}\;,\quad
\begin{cases}
	\lambda_i\\
	\gamma_i=-\varpi_i\\
	z_i=-\Omega_i
\end{cases}\;,
\end{equation}
where $\Omega_i$ is the $i$-th longitude of the ascending node. Then, we expand these expressions, neglecting the $\mathcal{O}(e_i^3)$ and $\mathcal{O}(s_i^3)$ remainders, and replace the result back into Eq.~\eqref{eqn:ham}. Subsequently, we apply the inverse transformations of the following Poincaré canonical coordinates in the same equation:
\begin{equation}
\label{eqn:Poinc}
\begin{cases}
		\Lambda_i\\
		x_i=\sqrt{2\Gamma_i}\cos\gamma_i\\
		u_i=\sqrt{2Z_i}\cos z_i\\
	\end{cases}\;,\quad
	\begin{cases}
		\lambda_i\\
		y_i=\sqrt{2\Gamma_i}\sin\gamma_i\\
		v_i=\sqrt{2Z_i}\sin z_i\\
	\end{cases}\;.
\end{equation}
Finally, we pass to the resonant canonical coordinates using
\begin{equation} 
\label{eqn:rescoord}
\begin{aligned}
&\begin{cases}
\sigma_1 = \lambda_1 \\
\sigma_2 = (p_1 + q_1)\lambda_3 - p_1 \lambda_2 \\
\sigma_3 = (p_2 + q_2)\lambda_4 - p_2 \lambda_3 \\
\sigma_4 = \lambda_4 \\
\sigma_5 = \lambda_5
\end{cases}\;,
\\
&\begin{cases}
\Sigma_1 = \Lambda_1 \\
\Sigma_2 = -\dfrac{\Lambda_2}{p_1} \\
\Sigma_3 = -\dfrac{\Lambda_3}{p_2} - \dfrac{p_1 + q_1}{p_1 p_2} \Lambda_2 \\
\Sigma_4 = \Lambda_4 + \dfrac{p_2 + q_2}{p_2} \Lambda_3 - \dfrac{(p_1 + q_1)(p_2 + q_2)}{p_1 p_2} \Lambda_2 \\
\Sigma_5 = \Lambda_5
\end{cases}\;.
\end{aligned}
\end{equation}
We note that the resonant combination $((p_1 + q_1)(p_2 + q_2)/g)\lambda_4-(p_1p_2/g)\lambda_2$ can be obtained as a linear combination of $\sigma_2$ and $\sigma_3$. Therefore, these are the only two independent canonical angles that survive the averaging procedure. As a consequence, the conjugate momenta $\Sigma_1$, $\Sigma_4$, and $\Sigma_5$ are constants of motion of the non-dissipative system. These constants will adiabatically drift over time when dissipation is included.

\subsection{Tidal effects}
\label{subsec:tides}

Among the various forces at play, tidal interactions are particularly significant. Tides introduce dissipation in the system through friction processes, in which orbital and rotational energy are exchanged between the planet and its satellites or dissipated as heat. In particular, moons within the system gain angular momentum from their hosting planet, resulting in an outward orbital migration; variations in their orbital separations can lead to encounters with mean motion resonances. At the same time, the planet's rotation rate $\omega_U$ decreases. This quantity can be computed throughout the dynamical evolution by exploiting the conservation of the total angular momentum of the system projected onto Uranus's spin axis:
\begin{equation}
    \label{eqn:angmom}
    L=C_U\omega_U+\sum_{i=1}^5\left(\beta_i\sqrt{\mu_ia_i(1-e_i^2)}+C_in_i\right)\cos I_i\;,
\end{equation}
which is equated to its current value. More precisely, in Eq.~\eqref{eqn:angmom} we assume that all bodies rotate about their axes of maximal inertia (gyroscopic approximation) and the moons are in 1:1 spin-orbit resonance with no obliquity. Here, $C_U=\zeta_Um_UR_U^2$, $C_i=\zeta_im_iR_i^2$ are the polar moments of inertia of Uranus and the moons, respectively, where $\zeta_U$, $\zeta_i$ are their normalized values dependent on the inner structure (\tablename~\ref{tab:param}). We will use Eq.~\eqref{eqn:angmom} in the context of tidal resonance locking, as clarified in Sect.~\ref{subsec:resloc}.

Dissipative forces generally do not fit within the Hamiltonian formalism, due to its inherently conservative nature. However, since the timescale of the dissipative effects is many orders of magnitude longer than the characteristic timescales of the conservative system (namely those of secular and resonant dynamics), we include tidal effects directly in the differential equations of motion as a small perturbation. Hence, we add to the Hamilton equations of the conservative averaged model secular terms due to tidal forces (up to second order in eccentricities and inclinations), which involve the evolution of the action-like orbital elements $a_i$, $e_i$, and $s_i$ (see \citealp{KAULA_1964,PEALE-etal_1979,MALHOTRA_1991,FERRAZ_MELLO-etal_2008}):
\begin{equation}
\label{eqn:tides}
\begin{aligned}
\frac{\dot{a}_i}{a_i}&=\frac23 
c_i\left(1 + \left(\frac{51}{4}-7D_i\right)e_i^2\right)\;,\\
\frac{\dot{e}_i}{e_i}&=-\frac13c_i\left(7D_i-\frac{19}{4}\right)\;,\\
\frac{\dot{s}_i}{s_i}&=-\frac16 c_i\;,
\end{aligned}
\end{equation}
with
\begin{equation}
\label{eqn:ciDi}
c_i=\frac92 \frac{k_{2,U}}{Q_{U,i}}\frac{m_i}{m_U}\left(\frac{R_U}{a_i}\right)^5n_i\;, \quad
D_i=\frac{k_{2,i}}{Q_i}\frac{Q_{U,i}}{k_{2,U}}\left(\frac{R_i}{R_U}\right)^5\left(\frac{m_U}{m_i}\right)^2\;.
\end{equation}
In the above, $R_i$ is the radius of the $i$-th satellite; $k_{2,U}$, $k_{2,i}$ are the tidal Love numbers of Uranus and the satellites, respectively; $Q_{U,i}$, $Q_i$ are the tidal quality factors of the planet (at the orbital frequency of the $i$-th moon) and of the $i$-th moon itself, respectively. These formulas hold as long as the central body rotates faster than the satellites' orbital motions (as in the case of the Uranus system), the satellites' rotation rates are synchronous with their orbital periods and the obliquities of the moons are zero.

The values of dissipative parameters of the Uranus system are largely unknown and probably varied during its billion-year history. Recent models set the Love number of Uranus $k_{2,U}=0.3$ \citep{STIXRUDE-etal_2021}, whilst past exploration of the moons' orbital evolution relied on an outdated value of $0.1$ (see \citealp{CUK-etal_2020,GOMES-CORREIA_2024a}). Traditional models give values of the quality factor of Uranus between $10\,000$ and $100\,000$ \citep{TITTEMORE-WISDOM_1988}, whereas recent works propose a much smaller value of $Q_U$ \citep{NIMMO_2023} and possibly dependent on the orbital frequency of the moons that generate the tides \citep{FULLER-etal_2016}. The tidal parameters of the satellites are even harder to constrain due to the limited knowledge of their interior structure and the lack of measurements. Moreover, orbital resonances can increase tidal heating, causing melting within the moons. In the case of icy satellites, the presence of subsurface oceans can greatly enhance the dissipative effects \citep{CASTILLO-etal_2023}. In our exploration, we focus on an epoch when tidal heating within Ariel was high because of the 2:1 resonance: we consider a dissipative parameter $k_{2,2}/Q_2$ of Ariel larger than those of the other moons. More precisely, we set $k_{2,2}=0.15$ and $k_{2,i}=0.03$ for all other moons. Moreover, we consider a low value of $Q_2$ between $50$ and $400$, so that $k_{2,2}/Q_2$ is between $3.75\times 10^{-4}$ and $3\times 10^{-3}$, while, similarly to \citet{CUK-etal_2020}, we set $Q_i=1\,000$ ($i=3,4,5$) and $Q_1=10\,000$ (\tablename~\ref{tab:param}). As a result, once outside the resonance, the damping in eccentricity of Ariel will be faster than that of the other satellites.

Finally, we express the set \eqref{eqn:tides} in terms of the canonical variables \eqref{eqn:Poinc} and \eqref{eqn:rescoord}, truncated at second order in eccentricities and inclinations. We then sum the result with the corresponding conservative parts (Hamilton equations) from Sect.~\ref{subsec:canvb} to obtain the final equations of motion.

\subsection{Tidal resonance locking}
\label{subsec:resloc}

Past studies of the evolution of the Uranian moons assumed constant or slightly variable values of the quality factor of the planet. However, a $Q_U$ as low as $1\,000$ \citep{NIMMO_2023,JACOBSON-PARK_2025} cannot be sustained for billions of years, as, in principle, inner moons would cover their distance from Uranus in far less time than their expected lifetime (see Eq.~\ref{eqn:tides}). Therefore, we must consider a tidal model that accounts for a significant variation in the value of $Q_U$.

The model that first predicted fast tidal migration of moons of giant planets is tidal resonance locking \citep{FULLER-etal_2016,LAINEY-etal_2020}. This theory states that tidal dissipation within planets depends strongly on the orbital frequency of the moon that generates tides. More precisely, if the orbital frequency matches the frequency of some interior oscillation mode within the planet, tidal dissipation at that frequency can increase to several orders of magnitude, enhancing the migration rate of the satellite as well. Furthermore, since the planet's interior evolves over time, mode frequencies change correspondingly, allowing this high-dissipation configuration to persist while the moon migrates at the rate dictated by the planet's interior evolution. \citet{FULLER-etal_2016} provide a formula for the variation of the quality factor of Uranus at the frequency of the $i$-th moon as a function of its semi-major axis:
\begin{equation}
\label{eq:Qreslock}
Q_{U,i}=\frac{9k_{2,U}}{2}\frac{m_i}{m_U}\left(\frac{R_U}{a_i}\right)^5\left(\frac{\omega_U}{n_i}-1\right)^{-1}n_it_{\alpha,i}\;,
\end{equation}
where $t_{\alpha,i}$ is the evolution timescale of the interior mode, which is generally different for the five satellites (see also \citealp{DOWNEY-etal_2020}). In particular, to get $Q_{U,i}=1\,000$ at the current epoch, it would be required that $t_{\alpha,i}\approx 2,\,3.3,\,53,\,1\,155,\,13\,500$ Gyrs, respectively. Although $\omega_U$ is expected to change slightly over time, we account for its evolution as described in Sect.~\ref{subsec:tides} (Eq.~\ref{eqn:angmom}).

Since the tidal locking mechanism depends on the coupling between the internal modes of the planet and the orbital frequency of the satellite, it is expected that not all moons are resonantly locked. In fact, it is possible that some moons did not match a mode frequency during their evolution, or, more probably, that they just experienced a temporary fast migration phase and then surpassed the mode frequency (see, e.g., \citealp{CUK-ELMOUTAMID_2023}). Therefore, in our setup, we can consider that during the evolution stage covered by our simulations some moons are in resonance locking (with an associated variable and currently low $Q_{U,i}$), and others are not (with an associated constant and medium-high $Q_{U,i}$). Specifically, in our simulations we set $Q_{U,i}=10\,000$ for moons not in resonance locking.

Following \citet{NIMMO_2023}, we always assume here that Ariel is in tidal resonance locking (see Sect.~\ref{sec:intro}). Therefore, in Eq.~\eqref{eqn:ciDi}, we consider that the quality factor of Uranus at the frequency of Ariel follows Eq.~\eqref{eq:Qreslock}.

Eq.~\eqref{eq:Qreslock} provides the variation of the quality factor in case the satellite is not in mean motion resonance with other moons. However, when Ariel is in resonance with Umbriel, it exchanges angular momentum with the outer moon, generally decreasing its own migration rate. As the orbital expansion of Ariel is dictated by the interior evolution of the planet, the resulting migration rate of Ariel must not change \citep{FULLER-etal_2016}. Therefore, while in mean motion resonance, the quality factor must be artificially reduced to obtain the actual quality factor. This is achieved by dividing $Q_{U,2}$ by a factor
\begin{equation}
\label{eq:facrl}
\mathcal{F}=1+\frac{m_3\sqrt{a_3}}{m_2\sqrt{a_2}}\;,
\end{equation}
which accounts both for keeping the tidal resonance lock of Ariel and pushing Umbriel to maintain the mean motion resonance. This factor is approximately $2.28$ when the two moons are locked in the 2:1 resonance. Eq.~\eqref{eq:facrl} corresponds to the factor dependent on the satellites' angular momentum ratio in Equation (19) of \citet{FULLER-etal_2016} that accounts for the extra dissipation.

\subsection{Numerical integration}
\label{subsec:numint}

The differential equations derived at the end of Sect.~\ref{subsec:tides} are obtained with the aid of symbolic computation software. These are converted into an operation-optimized form to minimize computational overhead and are then exported for numerical integration using a compiled programming language. All simulations are run forward using a Gauss-Radau integrator \citep{EVERHART_1985} modified using the tips of \citet{REIN-SPIEGEL_2015}.

We adopt Uranus-based units (mass of Uranus, radius of Uranus and years), so that, for example, $\mathcal{G}=3.4558207057809\times10^8~R_U^3m_U^{-1}\mathrm{yr}^{-2}$. The physical parameters and mean orbital elements of the Uranus system at the current epoch are listed in \tablename~\ref{tab:param}; references for these values are provided in the caption. In the numerical simulations, we vary the initial values of $a_i$, $e_i$, $I_i$, $\sigma_2$, and $\sigma_3$ depending on the type of resonance and statistical analysis being addressed (see Sects.~\ref{sec:2bres} and \ref{sec:break}).

When integration timespans are extremely large and multiple simulations are required, applying acceleration factors in presence of tidal dynamics is a common practice to speed up numerical integration (e.g., \citealp{TITTEMORE-WISDOM_1988,MALHOTRA_1991,LARI-etal_2020,GOMES-CORREIA_2024a}). This strategy is justified by the large timescale separation between the conservative and dissipative dynamics (Sect.~\ref{subsec:tides}). For the simulations of the 4:2:1 and 3:2:1 Ariel-Umbriel-Titania configurations (Sect.~\ref{sec:break}), we apply a $\times10$ acceleration factor. Remarkably, in the Uranus system, \citet{GOMES-CORREIA_2024a} obtained a reliable approximation of the real dynamics up to a $\times1000$ acceleration factor. Nevertheless, caution is required in the presence of three-body resonance crossings and tidal resonance locking, which motivates our choice. Accelerated simulations are carried out by dividing the quantities $Q_{U,i}$, $Q_i$ and $t_{\alpha,i}$ by the acceleration factor, along with the integration timespan, and then rescaling the resulting series by the same factor in time. This modification is purely artificial within the numerical implementation; all results and data reported in this work are ultimately restored to real-time values.

\begin{figure}
   \centering
   \includegraphics[scale=0.6]{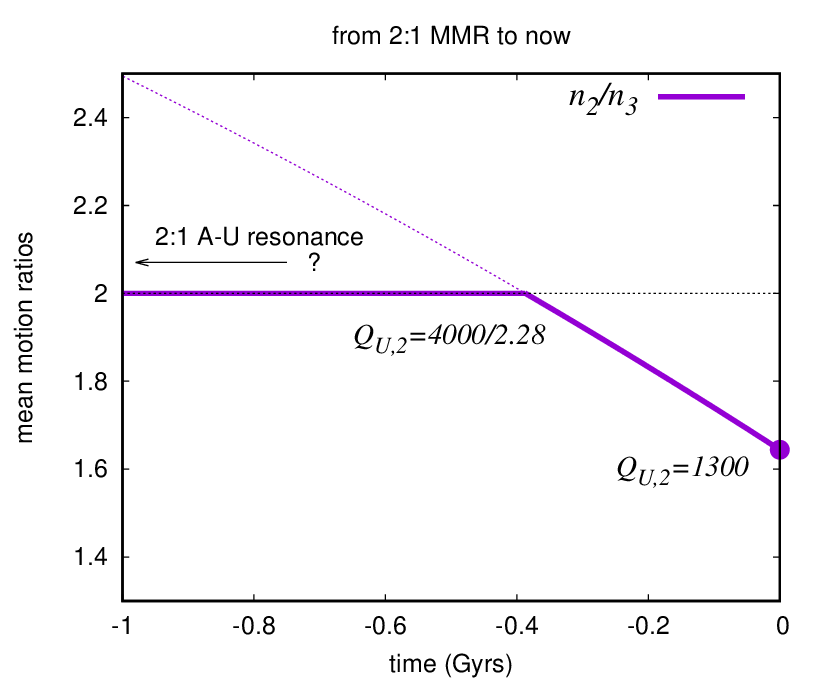}
   \caption{Secular variation of the mean motion ratio between Ariel and Umbriel (violet curve), obtained using $t_{\alpha,2}=4.2$ Gyrs and assuming that the moons were locked into the 2:1 resonance. The two values of $Q_{U,2}$ are computed from Eq.~\eqref{eq:Qreslock} at the epoch of the escape from the resonance and at the current epoch, respectively. During the mean motion resonance, we accounted for the extra dissipation factor $\mathcal{F}\approx2.28$. The violet dot indicates the current value of $n_2/n_3$, while the violet dashed line is the continuation of the evolution given by resonance locking without capture into the 2:1 resonance.}
   \label{fig:AUescape}
\end{figure}

\section{2:1 resonance between Ariel and Umbriel}
\label{sec:2bres}

If we set $Q_U\approx 1\,000$ and go back in time through Eq.~\eqref{eqn:tides} considering only the tides on the planet and disregarding the tidal evolution of the eccentricities and inclinations, the resonant crossing between Ariel and Umbriel is reached in just a few hundred million years (also disregarding any other resonant crossings; see, e.g., \citealp{PEALE_1986}). This is not necessarily the initial time of the resonance encounter between the moons, as the two satellites could have remained trapped into resonance for millions or even billions of years. However, it corresponds to the epoch when they escaped the 2:1 resonance, as depicted in \figurename~\ref{fig:AUescape}.

\begin{figure}
   \centering
   \includegraphics[scale=0.7]{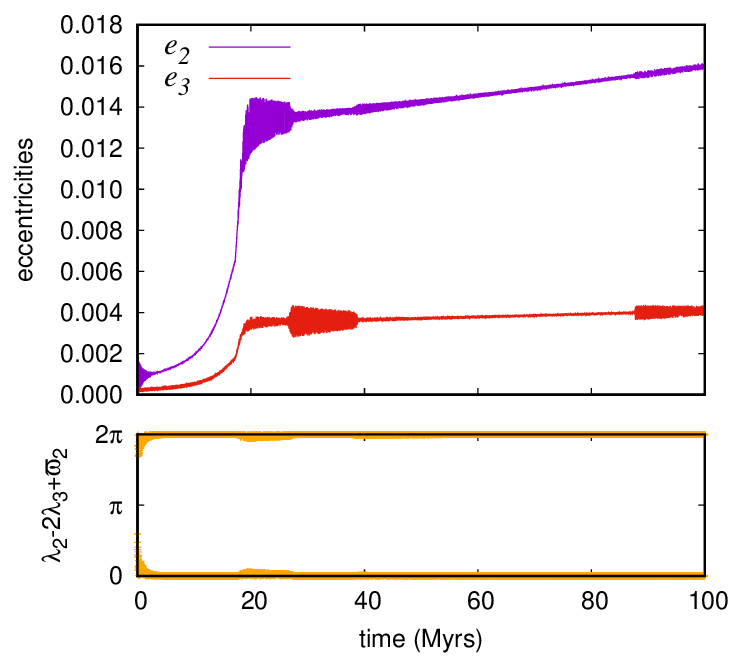}
   \caption{On the top, evolution of the eccentricities of Ariel and Umbriel during the capture into the 2:1 mean motion resonance. On the bottom, libration of the 2:1 resonant argument. This evolution has been obtained setting $t_{\alpha,2}=2.3$ Gyrs, $Q_2=100$ ($k_{2,2}=0.15$) and initial semi-major axes $a_2=6.150~R_U$ and $a_3=9.841~R_U$.}
   \label{fig:21capture}
\end{figure}

\begin{figure*}
   \centering
  \includegraphics[scale=0.7]{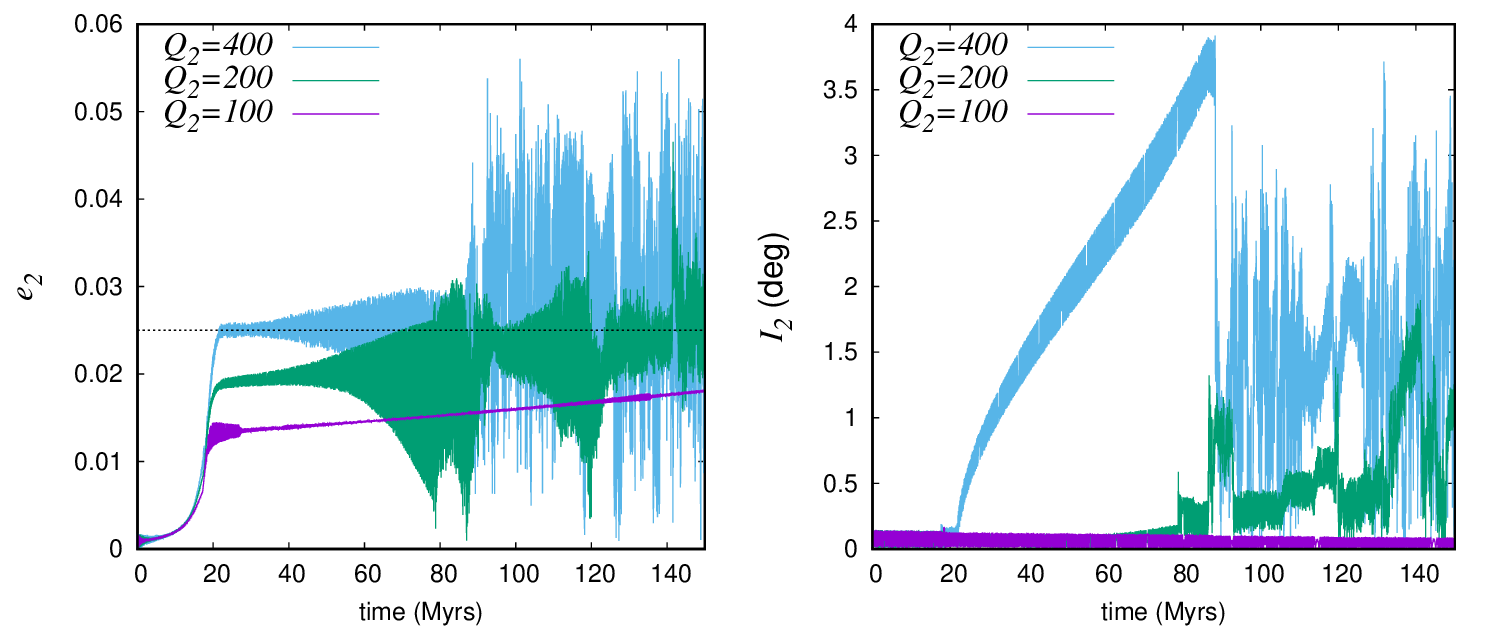}
   \caption{Evolution of the eccentricity and inclination of Ariel during the capture into the 2:1 mean motion resonance with Umbriel, considering different values of the parameter $Q_2$ ($k_{2,2}=0.15$). The dotted line represents the eccentricity threshold (see text) above which the inclination resonance is triggered. These simulations have been obtained setting $t_{\alpha,2}=2.3$ Gyrs and initial semi-major axes $a_2=6.150~R_U$ and $a_3=9.841~R_U$.}
   \label{fig:ei2Q2}
\end{figure*}

In order to determine the effects of the resonance, we set Ariel's and Umbriel's semi-major axes such that $n_2/n_3 \gtrsim 2$ and run numerical simulations of the 2:1 resonant encounter with the dynamical averaged model presented in Sect.~\ref{sec:dynmo}. We consider that Ariel is in resonance locking and that it had a convergent resonant encounter with Umbriel (i.e., $n_2/n_3$ decreased). This is accomplished if Umbriel migrated slower than Ariel. In the simulations, we consider that the quality factor of Uranus at the frequency of Umbriel is constant and equal to $10\,000$. This is motivated by the fact that, once captured, Umbriel would be pushed by Ariel, so it would anyway lose its potential resonance lock with an interior mode of Uranus. Although $t_{\alpha,2}$ for Ariel is not known, we can use the predictions of \citet{NIMMO_2023} on the value of the Uranus's quality factor during the lock into the 2:1 mean motion resonance to set limits on its value, i.e., $1\,000\lesssim Q_{U,2}\lesssim 2\,500$. Using Eq.~\eqref{eq:Qreslock} and considering the enhancing factor $\mathcal{F}$ (Eq.~\ref{eq:facrl}), we obtain $2.3\le t_{\alpha,2}\le 6.0$ Gyrs, which places the epoch of the escape from the resonance between about $200$ and $600$ Myrs ago. Interestingly, setting $t_{\alpha,2}=2.3$ Gyrs results in a current value of $Q_{U,2}$ of $698$, which is very close to the value estimated by \citet{JACOBSON-PARK_2025} for the quality factor of Uranus. The given range for $t_{\alpha,2}$ corresponds to a rate in semi-major axis $\dot a_2$ between $8$ and $21$ cm/yr, which is of the same order of those measured for Ganymede and Titan \citep{LAINEY-etal_2009,LAINEY-etal_2020}

\begin{figure}
   \centering
   \includegraphics[scale=0.7]{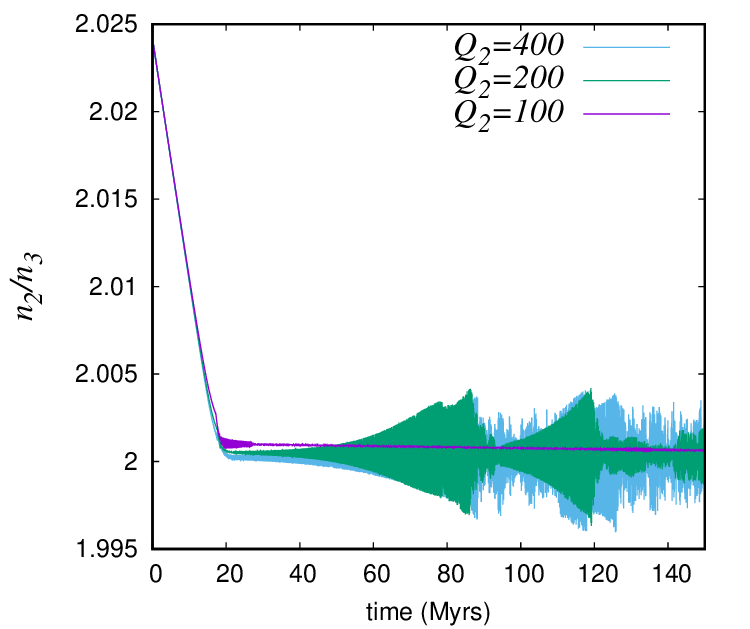}
   \caption{Evolution of the ratio between the mean motions of Ariel and Umbriel during the capture into the 2:1 mean motion resonance, considering different values of the parameter $Q_2$ ($k_{2,2}=0.15$).}
   \label{fig:n2n3Q2}
\end{figure}

Once we set $t_{\alpha,2}$, the initial epoch of the resonant encounter depends on the initial semi-major axis of Umbriel, which is unknown. In order to keep the integration time low and give time to the orbits to relax after the resonance capture, we select an epoch about $100$ Myrs before the expected escape. More precisely, we start from $a_2=6.150~R_U$ and $a_3=9.841~R_U$. At this stage of the study, this choice is arbitrary. Choosing smaller values of the semi-major axes would set the resonant encounter earlier in history and would require the lock into mean motion resonance to last longer (possibly even billions of years).

For these simulations, we set the initial eccentricities of the moons to very small values ($e_i = 10^{-4}$), assuming that they have been damped by tidal dissipation, while we set the initial inclinations, longitudes and remaining semi-major axes to their current mean values (\tablename~\ref{tab:param}). In addition, as we start far enough from the resonance, we arbitrarily set the initial value of the resonant variable $\sigma_2$. As observed by \citet{TITTEMORE-WISDOM_1990} and \citet{CUK-etal_2020}, resonant crossings occurring after the 2:1 Ariel-Umbriel mean motion resonance could have been responsible for the increase in Miranda's inclination. We address this possibility in Sect.~\ref{sec:break}, where we set $I_1\approx0$. 

Starting from such small eccentricities, we obtain that the capture into the 2:1 mean motion resonance is certain, as confirmed by all the simulations performed for different values of $k_{2,2}/Q_2$ and expected by classical adiabatic evolution (see, e.g., \citealp{MURRAY-DERMOTT_2000}). As the 2:1 resonant angle starts to librate, the eccentricities of Ariel and Umbriel increase toward their equilibrium values, set by the balance between resonant and tidal effects (see \figurename~\ref{fig:21capture}). When Ariel's eccentricity reaches its equilibrium value, the tidal heating within the moon is maximum. In the simulation reported in \figurename~\ref{fig:21capture} and obtained with $k_{2,2}/Q_2=1.5\times10^{-3}$, we get $e_2\approx 0.014$, which induces a tidal heating of about $250$ GW (see Eq.~\ref{eq:endis} in Sect.~\ref{subsec:therm}). This is the magnitude expected by \citet{NIMMO_2023} for $Q_{U,2}\approx 1\,000$ and that can account for the past heat flux predicted for Ariel \citep{PETERSON-etal_2015}. We observe that in these simulations the mean values of the eccentricities do not remain perfectly constant, but increase slowly because of the change of $Q_{U,2}$ given by Eq.~\eqref{eq:Qreslock}.

It is worth noting that this result is not limited to the very small initial $e_i$ we chose. In fact, for not being captured into resonance, quite high pre-resonance values of the eccentricities would be required, at least larger than $0.01$ (see \citealp{TITTEMORE-WISDOM_1990}), which seems very unlikely to be attained before the resonant encounter. Therefore, if the 2:1 resonance was crossed, Ariel and Umbriel would have been almost certainly captured.

When considering different values of $k_{2,2}/Q_2$ (fixed $k_{2,2}$ and variable $Q_2$), we obtain different values of the equilibrium eccentricities, as the balance between tidal damping and resonance pumping is reached at different $e_2$. \citet{TITTEMORE-WISDOM_1990} provided the following analytical formula for the equilibrium eccentricity of Ariel:
\begin{equation}
\label{eq:e2eq}
    e_{2,\text{eq}}=\sqrt{\frac{1}{14D_2\left(1+2\displaystyle\frac{m_2}{m_3}\left(\frac{a_2}{a_3}\right)^2\right)}}\;,
\end{equation}
which matches the results of the numerical simulations.

However, we note that if the value of $Q_2$ sets an equilibrium eccentricity larger than a certain threshold (in the configuration used for simulations in \figurename~\ref{fig:ei2Q2}, this threshold is approximately $0.025$), Ariel's increase in eccentricity stops at this threshold value, and the moons trigger an inclination resonance that drives a substantial growth of Ariel’s inclination, as shown in \figurename~\ref{fig:ei2Q2} in the case $Q_2=400$. Depending on the type of inclination resonance, an increase in inclination can also affect Umbriel. This was already observed by \citet{TITTEMORE-WISDOM_1990}, who found that this kind of evolution appears in the eccentric-inclined model, but not in the planar model where inclinations are neglected. As the inclinations can increase up to few degrees while the moons are in resonance, and are difficult to damp after the exit, we shall try to avoid these kinds of evolution in reconstructing the moons' history. Considering $Q_{U,2}\approx 1\,000$, we obtain that the equilibrium eccentricity reaches the threshold of $0.025$ for $Q_2\approx 250$. In the following section, we will take into account this upper limit when exploring the evolution of the moons through and after the 2:1 resonance.

Furthermore, for values of $Q_2\ge 200$, we observe an excitation of the resonant state of the moons which increases the oscillation amplitude of the eccentricities around their forced values, as shown in \figurename~\ref{fig:ei2Q2} in the case $Q_2=200$ and $Q_2=400$. This can produce overlapping between eccentricity and inclination resonances, resulting in a chaotic evolution of the orbital elements. Therefore, choosing small values of $Q_2$ while Ariel is in resonance seems to be preferable, as the inclinations of the moons are not affected and remain small (see \figurename~\ref{fig:ei2Q2} with $Q_2=100$).

It is worth noting that the effects described in this section depend on the ratio between $Q_2$ and $Q_{U,2}$ (see Eqs.~\ref{eqn:ciDi} and~\ref{eq:e2eq}). Therefore, if we set a different value of $t_{\alpha,2}$, the same results hold with a rescaled value of the quality factor of Ariel.

Whatever the value of $Q_2$ and the kind of evolution (eccentricity or inclination resonances, regular or chaotic motion), in all our simulations we observe that the 2:1 mean motion resonance between Ariel and Umbriel persists once the two moons are captured (see \figurename~\ref{fig:n2n3Q2}). This seems to be in contradiction with the current configuration of the system, and this is the reason why the 2:1 resonance was excluded from the orbital history of the moons \citep{TITTEMORE-WISDOM_1990}. In the next section, we will describe a dynamical mechanism that could have disrupted the 2:1 resonance between Ariel and Umbriel.

\section{Breaking the 2:1 resonance}
\label{sec:break}

The 2:1 mean motion resonance between Ariel and Umbriel is probably the best candidate to explain the resurfacing of Ariel, and the fast tidal migration scenario offers a new opportunity to include such a resonance in the history of the Uranian moons.

\begin{figure*}
   \centering
   \includegraphics[scale=0.65]{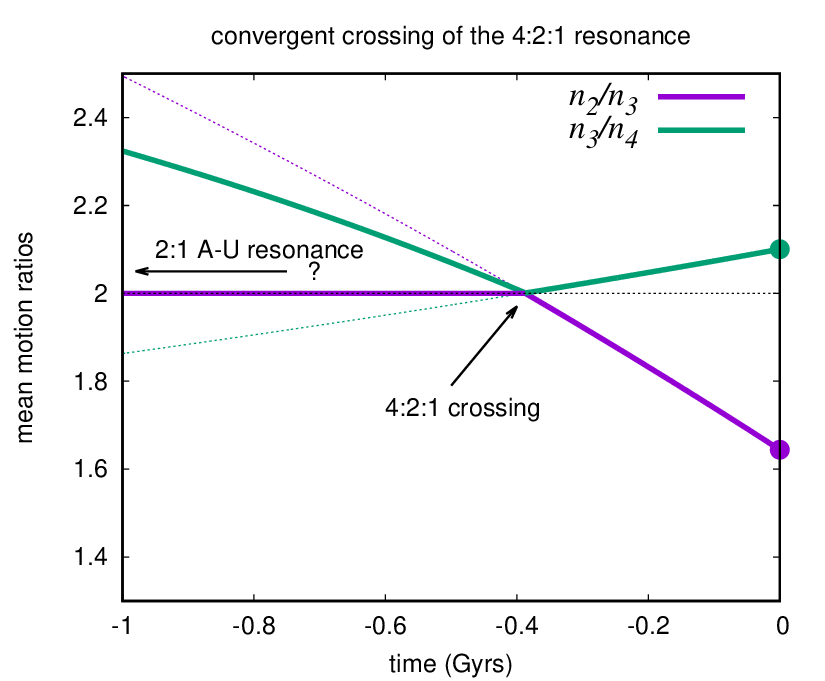}\includegraphics[scale=0.65]{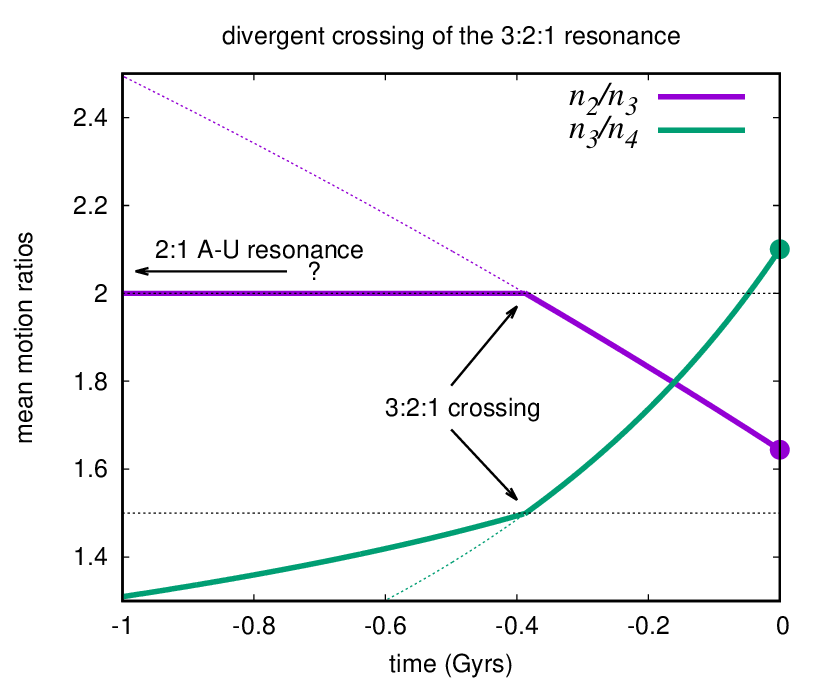}
   \caption{Secular variation of the mean motion ratio of Ariel-Umbriel (violet curve) and Umbriel-Titania (green curve) in the two scenarios considered in Sect.~\ref{sec:break}, obtained using $t_{\alpha,2}=4.2$~Gyrs and $t_{\alpha,4}=80$~Gyrs (left) or $10.4$~Gyrs (right). The plots are computed similarly as in Sect.~\ref{sec:2bres} (\figurename~\ref{fig:AUescape}). The two evolutions assume that the two inner moons were locked into the 2:1 mean motion resonance and the crossing of the three-body resonant chain 4:2:1 (left) or 3:2:1 (right) with Titania succeeded in breaking the resonance. The violet and green dots indicate the current values of $n_2/n_3$ and $n_3/n_4$, respectively, while the violet and green dashed lines are the continuation of the evolution given by resonance locking without capture into the 2:1 resonance.}
   \label{fig:AUTescape}
\end{figure*}

So far, no authors have managed to find a way to break the resonance \citep{DERMOTT-etal_1988,PEALE_1988,TITTEMORE-WISDOM_1990}. \citet{PEALE_1988} conjectured that the perturbation of Titania could have played a role in the disruption of the resonance. However, the secular perturbation of Titania is not strong enough for this task. Nonetheless, a resonant interaction with Titania could be the key to breaking the 2:1 resonance. Such a resonant encounter would result in a three-body resonant interaction between the three moons.

As currently $n_2/n_4\approx 3.452$ and $n_3/n_4\approx 2.100$, the closest strong three-body resonances that the three satellites could have encountered are the 4:2:1 (Laplace) and 3:2:1 (Laplace-like). As we assume that Ariel and Umbriel were already in the 2:1 resonance, these resonant chains involve the first-order two-body resonances 2:1 and 3:2, respectively, between Umbriel and Titania. It is worth noting that also \citet{CUK-etal_2020} investigated the dynamical effects on the system due to a three-body resonance among these three moons, which in their case was a combination between the 5:3 Ariel-Umbriel and the 2:1 Umbriel-Titania resonances.

Before exploring the outcome of these three-body resonant encounters through numerical simulations, we sketch the expected secular evolution after the potential breaking of the 2:1 resonance. While $n_2/n_3$ decreases from $2$ to its current value ($1.644$), the ratio $n_3/n_4$ must increase from $2$ (or $1.5$) to its current value ($2.100$), as depicted in \figurename~\ref{fig:AUTescape}. Therefore, we require that, after the breaking of the resonance, Titania migrates faster than Umbriel, even though Umbriel is closer to Uranus than Titania. This provides constraints on the migration rate of Titania and the tidal quality factor of Uranus at the frequency of this moon. More precisely, it is required a value of $Q_{U,4}$ of about $100$ or even lower. Such a small value implies that $Q_{U,4}$ must vary significantly during the evolution, for example following a resonance locking formulation (see Eq.~\ref{eq:Qreslock}), as already assumed for Ariel.

Since we consider a value of $t_{\alpha,2}$ between $2.3$ and $6.0$ Gyrs, it follows that $t_{\alpha,4}$ must lie between approximately $45$ and $110$ Gyrs for the scenario involving the 4:2:1 resonant chain, and between approximately $5.7$ and $15$ Gyrs for the 3:2:1 resonant chain. This sets the current value of $Q_{U,4}$ from $40$ to $100$ ($\dot a_4$ between $4$ and $10$ cm/yr) and from $5$ to $13$ ($\dot a_4$ between $30$ and $80$ cm/yr), respectively, which is one or two orders of magnitude smaller than the one supposed for Ariel. Such values are in line with what was measured for Titan in the Saturn system and proposed for Callisto in the Jovian system \citep{FULLER-etal_2016,LAINEY-etal_2020,LARI-etal_2023}. In the first scenario (4:2:1 resonance) Ariel migrates faster than Titania, while in the second one (3:2:1 resonance) Titania migrates faster than Ariel (cf. Sect.~\ref{sec:2bres}).

Assuming tidal resonance locking for Titania and not for Umbriel is coherent with the evolution depicted in \figurename~\ref{fig:AUTescape}. Indeed, as already observed, once captured into the 2:1 resonance, Umbriel orbital frequency would have been pushed away from the driving interior mode of Uranus by the faster migration rate provided by Ariel. Therefore, once the moons exit from the mean motion resonance, it is reasonable to assume that Ariel and Titania conserve their tidal resonant lock, while Umbriel does not.

In the following, we present numerical simulations of these triple resonant encounters obtained with the dynamical averaged model presented in Sect.~\ref{sec:dynmo}. We draw statistics of the outcome, focusing on the evolution for which the 2:1 mean motion resonance between Ariel and Umbriel is disrupted.

\subsection{Crossing of 4:2:1 resonance with Titania}
\label{subsec:421res}

We run numerical simulations of the crossing of the 4:2:1 resonant chain between Ariel, Umbriel and Titania while the first two moons are locked into the 2:1 mean motion resonance. In order to track the full process of capture and (hopefully) escape from the resonance, we set initial semi-major axes $a_2=6.150~R_U$ and $a_3=9.841~R_U$ as in Sect.~\ref{sec:2bres}. Moreover, we start from $a_4=16.195~R_U$, in order to initially have $n_3/n_4>2$. We take the lowest values of $t_{\alpha,2}$ and $t_{\alpha,4}$ in order to get the fastest admissible evolution and limit the computation time. More precisely, we set $t_{\alpha,2}=2.3$ Gyrs and $t_{\alpha,4}=45$ Gyrs, so that the three-body resonance crossing occurs about $100$ Myrs after the capture into the 2:1 mean motion resonance between Ariel and Umbriel, that is, after the resonance equilibrium eccentricity is achieved.

\begin{figure}
\centering
\includegraphics[scale=0.205]{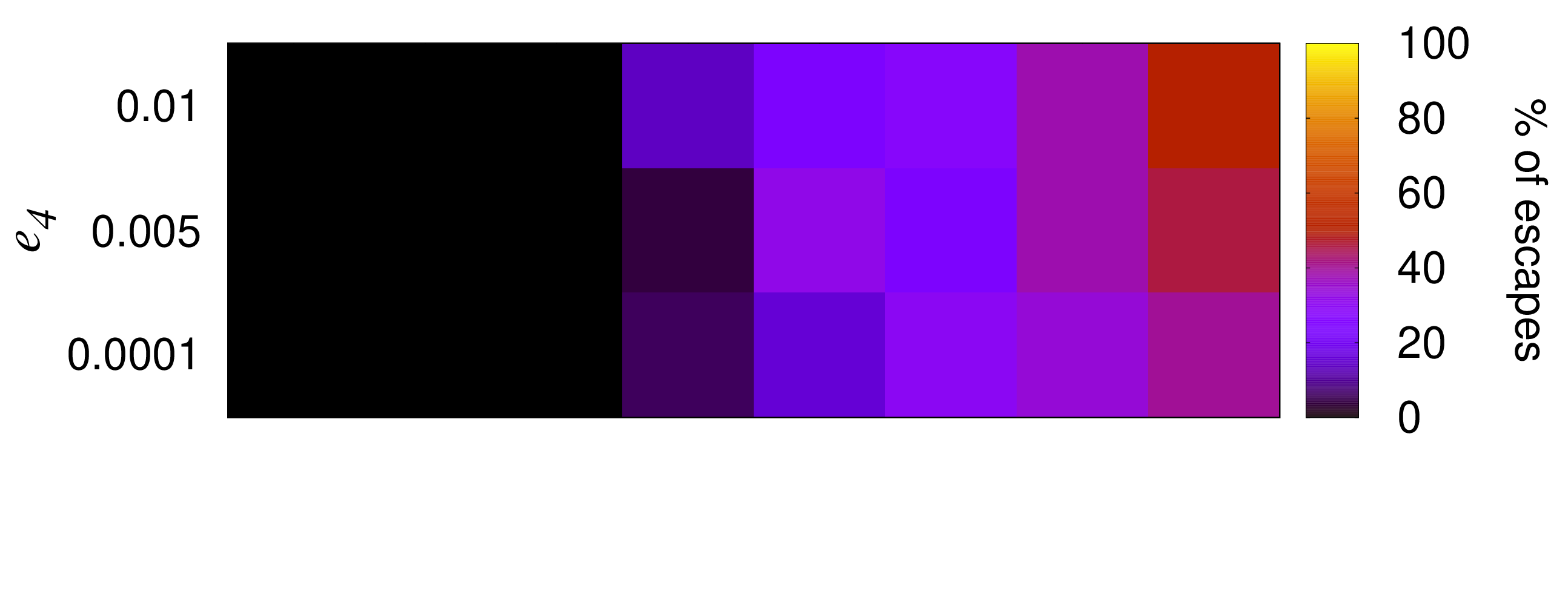}\\
\vspace{-9mm}
\includegraphics[scale=0.205]{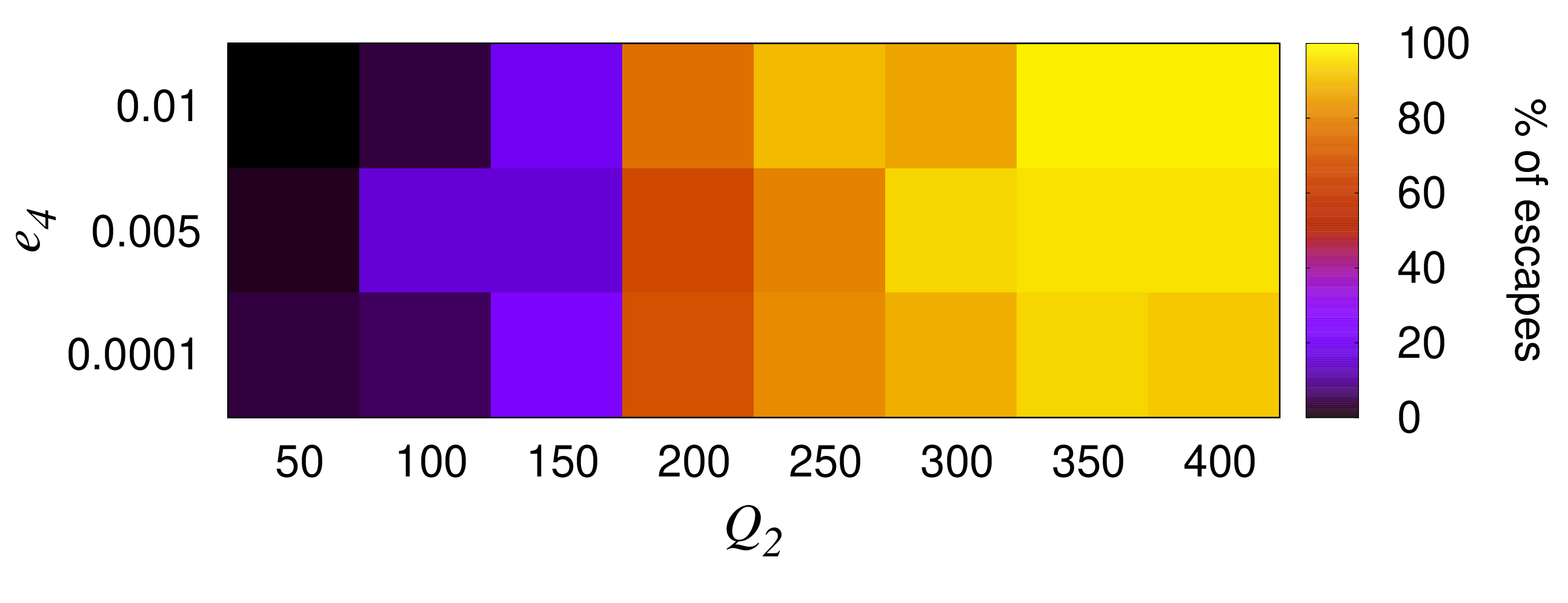}
\caption{Percentage of simulations for which the 2:1 mean motion resonance between Ariel and Umbriel breaks down per every setup (each setup counts a total of 50 simulations). On the top, simulations on the 4:2:1 resonant crossing; on the bottom, simulations on the 3:2:1 resonant crossing.}
\label{fig:escapes}
\end{figure}

Also for these simulations we start with small eccentricities of all moons ($e_i=10^{-4}$), even though we additionally explore cases where $e_4$ is slightly higher, in order to check its impact during the three-body resonant crossing. Initial inclinations are set to the same values of today (\tablename~\ref{tab:param}), except for Miranda, where we set a very low value ($I_1=10^{-4}$ rad), in order to investigate whether the proposed scenario could have contributed to its current high value.

We run $50$ simulations per setup, varying only the initial value of $\sigma_2$ within the same setup, which is uniformly sampled between $[0,2\pi)$ radians; the initial value of the second resonant variable $\sigma_3$ is chosen arbitrarily. The setups differ for values of $Q_2$ (fixing $k_{2,2}=0.15$), that we take between $50$ and $400$ with spacing $50$, and initial $e_4$, chosen to be $0.0001$, $0.005$ and $0.01$. The total propagation time is $150$ Myrs, sufficient for checking the possible disruption of the 2:1 resonance and the values of the orbital elements after the three-body resonance crossing.

We divide the simulations according to the outcome of the resonant encounter with Titania: the 2:1 mean motion resonance between Ariel and Umbriel breaks down (case A) or it survives (case B). We can immediately notice from \figurename~\ref{fig:escapes} (top panel) that the simulations of case A are less than those of case B in every setup we considered. Moreover, no escapes are found for $Q_2\le 150$, and just a few for $Q_2=200$. This implies that for disrupting the 2:1 resonance through the 4:2:1 three-body resonant encounter, a significant excitation of the moons' eccentricities and inclinations is necessary (see, e.g., the evolution for $Q_2=200$ or $400$ in \figurename~\ref{fig:ei2Q2}). For higher values of $Q_2$ (i.e., less dissipation in Ariel), the escape probability increases up to about $50\%$. The initial value of $e_4$ does not appear to be as determinant to the breaking of the resonance as the quality factor of Ariel, although a mild trend becomes noticeable as $e_4$ increases for higher values of $Q_2$.

\begin{figure*}
   \centering
   \includegraphics[scale=0.45]{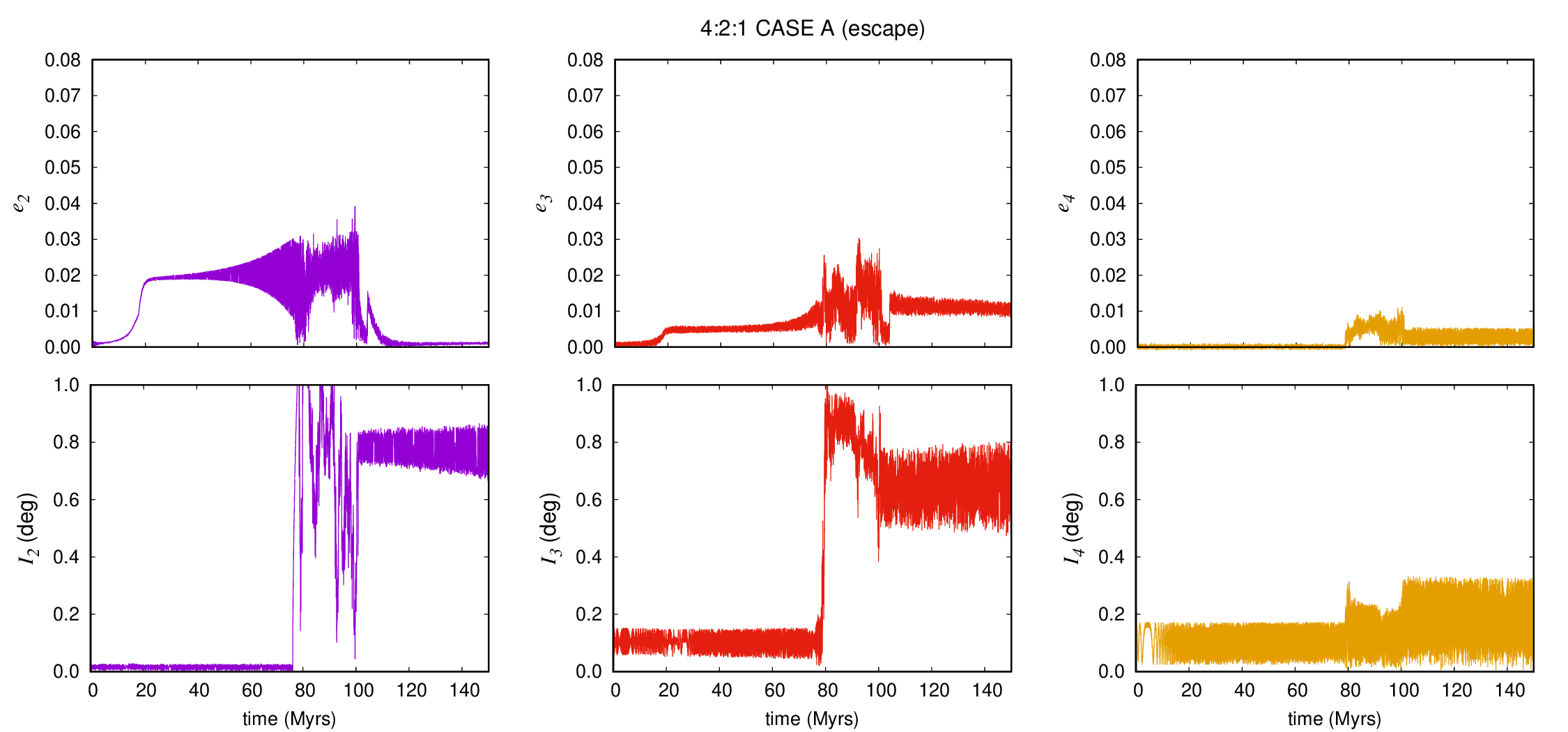}
   \includegraphics[scale=0.45]{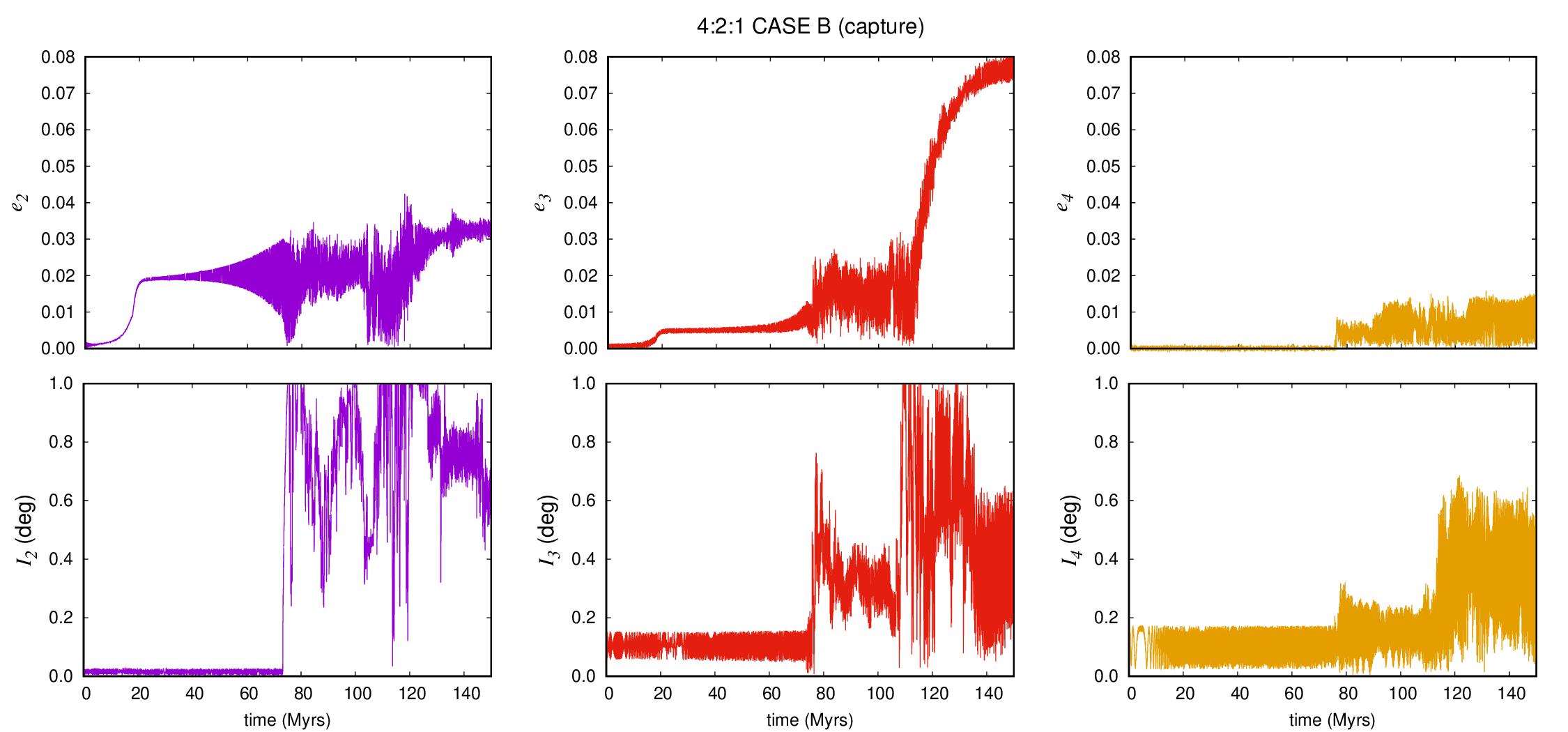}
   \includegraphics[scale=0.65]{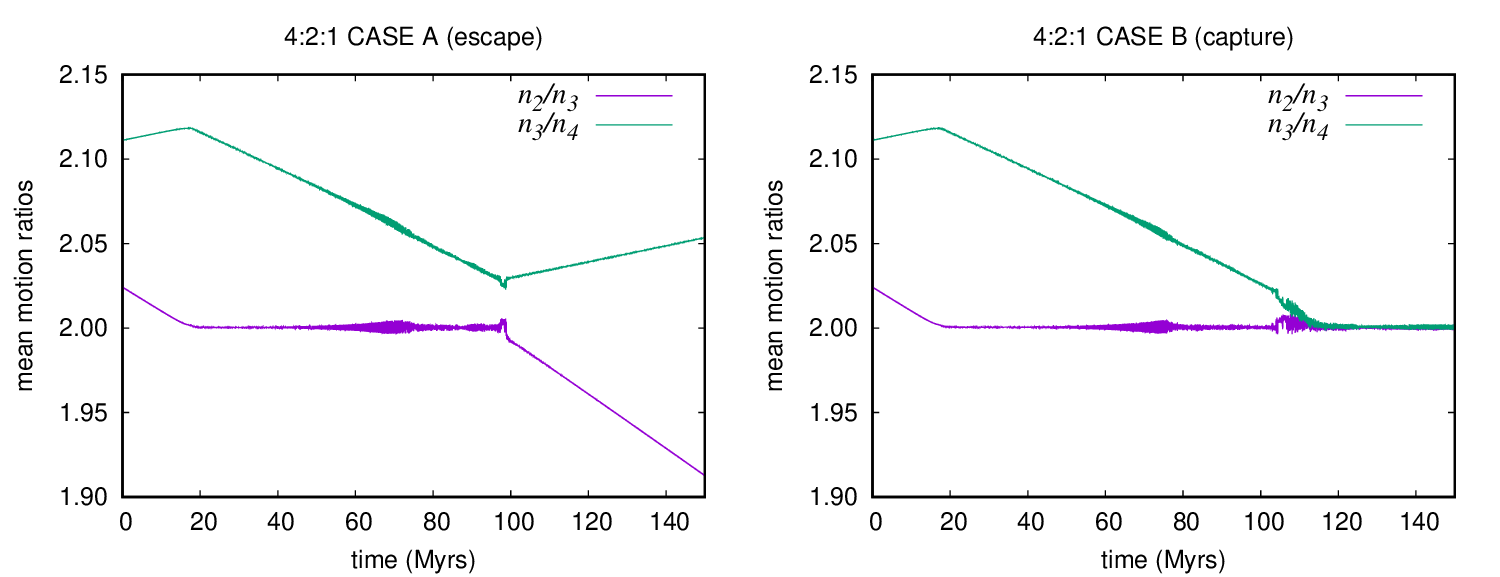}
   \caption{Simulations of the crossing of the 4:2:1 resonant chain between Ariel, Umbriel and Titania, obtained setting $t_{\alpha,2}=2.3$ Gyrs, $Q_2=200$ and initial $e_4=0.0001$. On the top, evolution of the eccentricities and inclinations of the three moons in a case where the 2:1 mean motion resonance between Ariel and Umbriel is disrupted (case A). In the middle,  evolution of the eccentricities and inclinations of the three moons in a case where also Titania is captured into mean motion resonance with Ariel and Umbriel (case B). On the bottom, evolution of the ratios of the mean motions in both cases.}
   \label{fig:421cases}
\end{figure*}

All simulations of case B end up in the formation of a three-body Laplace resonance between Ariel, Umbriel and Titania. In \figurename~\ref{fig:421cases} (middle plots), we report the evolution of the orbital elements of the three satellites obtained in one of the simulations with $Q_2=200$ and initial $e_4=0.0001$. We observe that, due to the chaotic behavior triggered during the lock into the 2:1 resonance, all moons' eccentricities and inclinations vary wildly before the three-body resonant encounter. Once Titania is captured, forming the three-body resonance, the eccentricities are forced to new high values by the new resonant relation. Since the system remains trapped into resonance, we consider this kind of evolution not compatible with the current orbital configuration of the Uranian system.

Simulations of case A are more interesting for our analysis, as they prove that the resonant encounter with Titania can succeed in disrupting the strong 2:1 resonance between Ariel and Umbriel. In \figurename~\ref{fig:421cases} (top plots), the reported evolution is obtained in the same setup with $Q_2=200$ and initial $e_4=0.0001$. The approach to the three-body resonance is similar to the previous case. However, in this case, we observe from \figurename~\ref{fig:421cases} (bottom plot on the left) that, as Titania enters into the resonant region, the 2:1 resonance is disrupted and the mean motion ratio $n_2/n_3$ decreases, while $n_3/n_4$ increases. Once the resonance breaks down and the moons move away, the evolution becomes stable, with eccentricities and inclinations that do not vary much anymore, apart from $e_2$, which is quickly damped because of the relatively low value of $Q_2$.

\begin{figure*}
\centering
\includegraphics[scale=0.475]{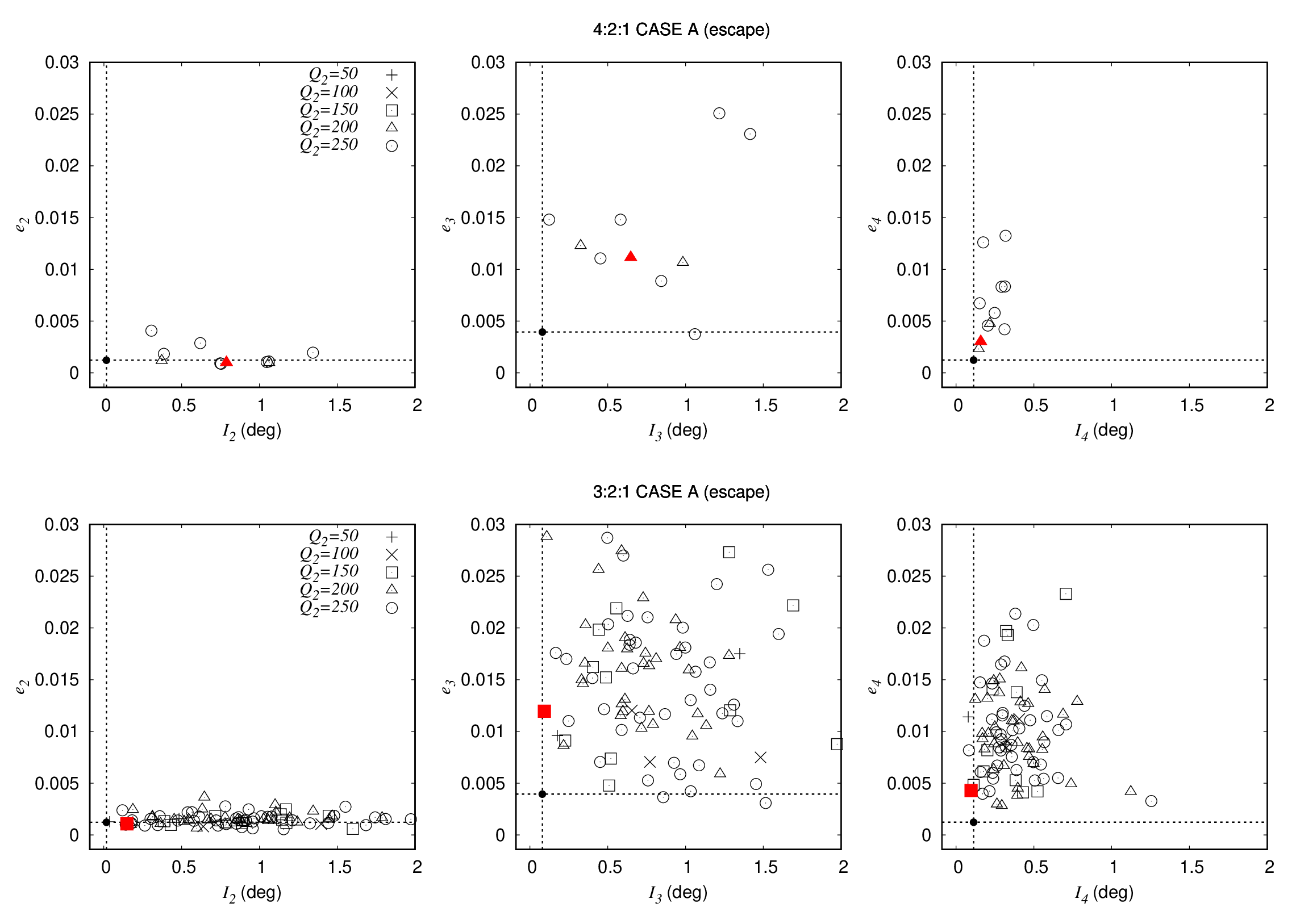}
\caption{Mean orbital eccentricities and inclinations of Ariel, Umbriel and Titania obtained from the numerical simulations (only the ones with initial $e_4=0.0001$ and $Q_2\le 250$) after the disruption of the 2:1 mean motion resonance, both in the case of 4:2:1 resonant crossing (top) and 3:2:1 resonant crossing (bottom). The values are taken when the system reached $n_2/n_3=1.94$ (see text). The dashed lines indicate the current values of the mean orbital elements. The red points show the orbital elements of the simulations presented in \figurename~\ref{fig:421cases} and~\ref{fig:321cases}.}
\label{fig:incecc}
\end{figure*}

The disruption of the resonance does not occur at the nominal 4:2:1 ratio of the mean motions, but slightly before. This is because the overlap of 2:1 resonances (Ariel-Umbriel and Umbriel-Titania) generates a jungle of three-body resonances that surrounds the nominal location of the 4:2:1 resonance (see \citealp{LARI-etal_2020} for details). Moreover, after the breaking of the 2:1 resonance, there is a last kick on the eccentricities of Ariel and Umbriel. The kick can form because of the crossing of one of the remaining resonant combinations, or even the nominal 4:2:1. In the subsequent evolution, eccentricities can then be damped by tidal dissipation acting within satellites (for instance as a result of partial melting in the interior of the moons), especially if the values of $k_{2,i}/Q_i$ are large, as in the case of Ariel. 

Although the simulations within case A eventually manage to reach a final non-resonant configuration, they do not generally fit exactly the current orbital configuration of the system. In fact, because of their chaotic variation before the escape from the resonance, the orbital elements end up to final values that vary stochastically, with no preferences to very low values (see \figurename~\ref{fig:incecc}, top panel). In particular, the values of inclinations at the exit are generally larger than their current ones. Final eccentricities and inclinations are computed once Ariel and Umbriel have sufficiently passed beyond the 2:1 resonance, prior to any subsequent (possible) resonant crossings. For this purpose, we adopt $n_2/n_3 = 1.94$ as the appropriate threshold in all simulations involved.

In the end, the scenario involving the crossing of the 4:2:1 can succeed in breaking the resonance. However, the probability of reproducing the current configuration of the moons is low, unless further dynamical mechanisms affected the orbital elements in the evolution following the breaking of the 2:1 resonance (see the discussion in Sect.~\ref{sec:discu}).

\begin{figure*}
   \centering
   \includegraphics[scale=0.45]{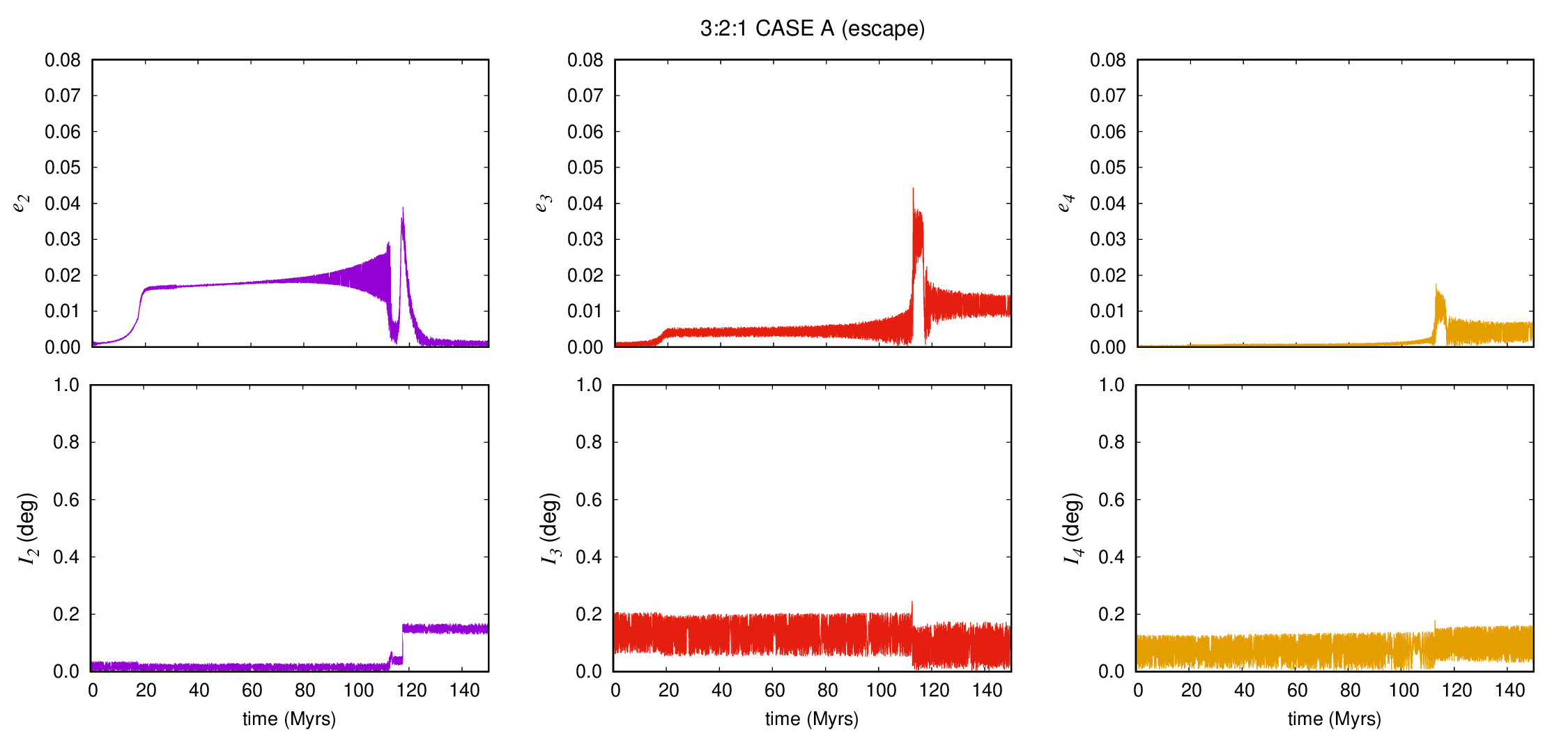}
   \includegraphics[scale=0.45]{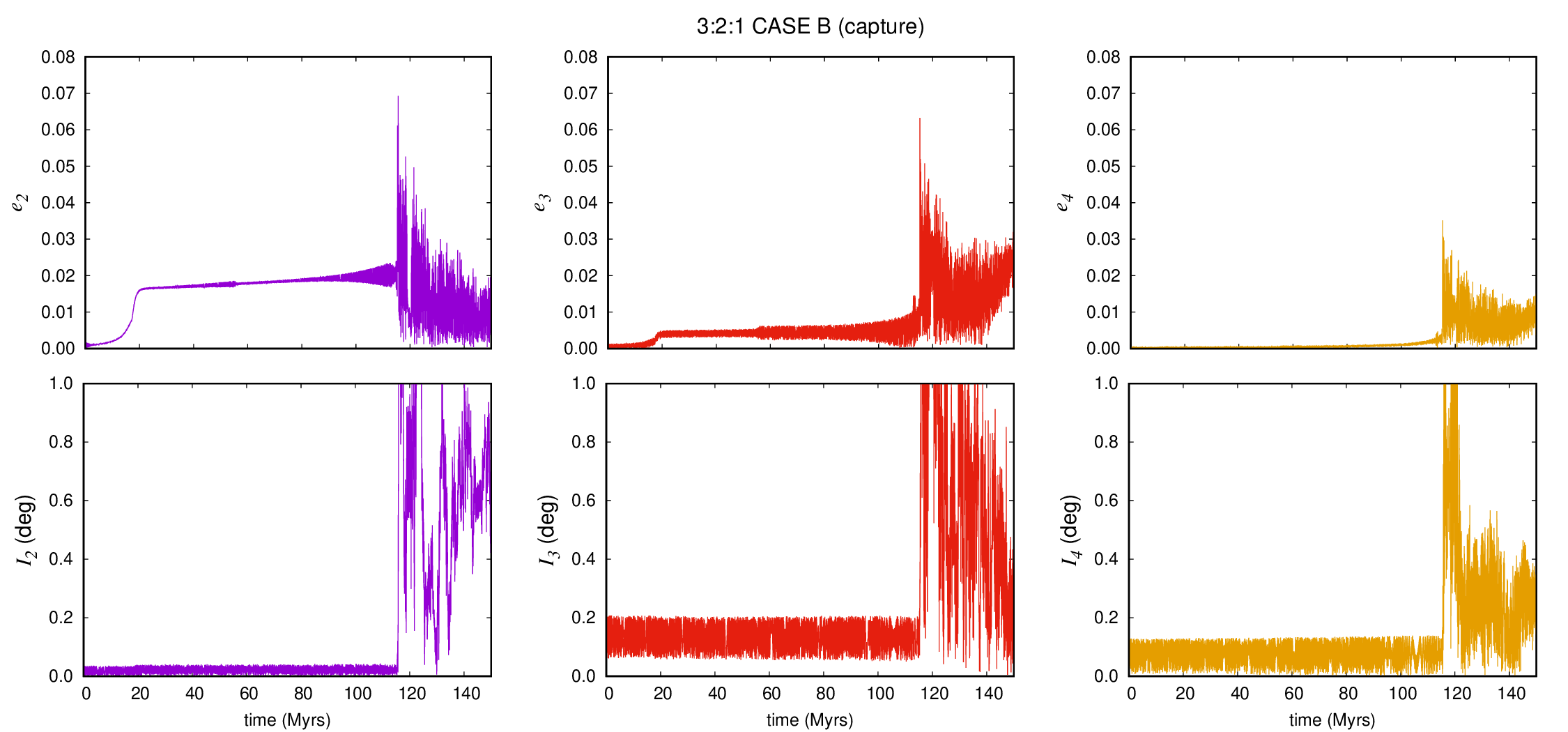}
   \includegraphics[scale=0.65]{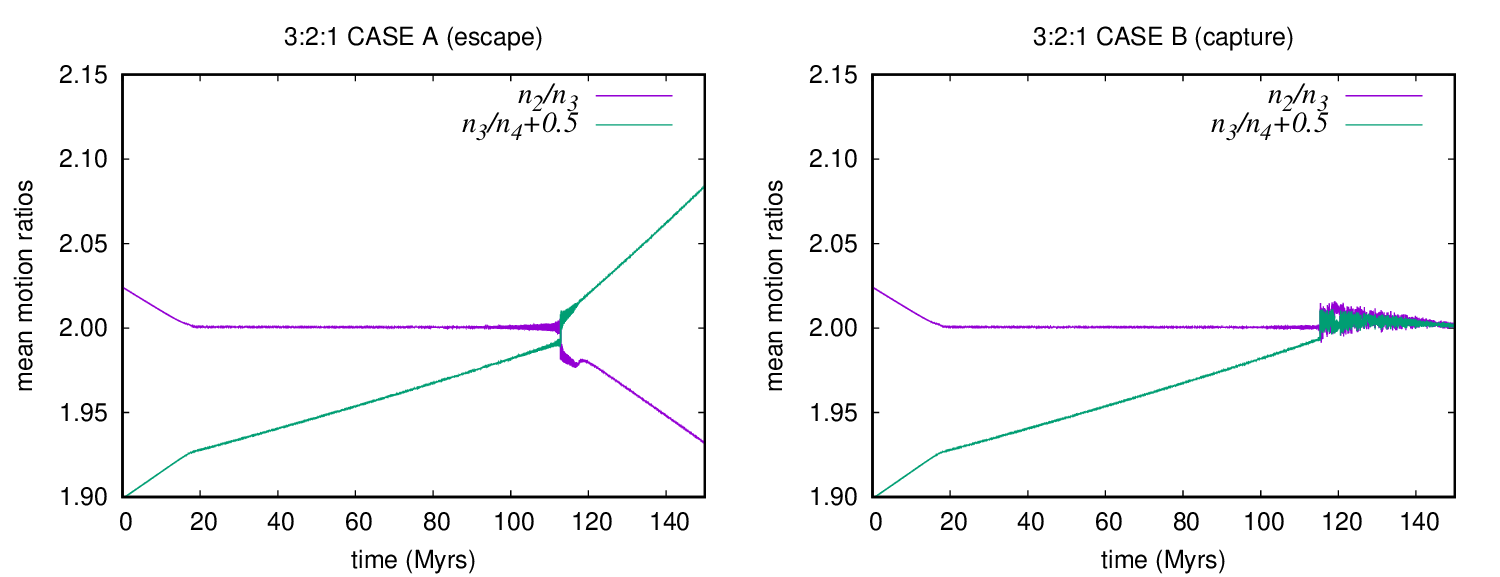}
   \caption{Simulations of the crossing of the 3:2:1 resonant chain between Ariel, Umbriel and Miranda, obtained setting $t_{\alpha,2}=2.3$ Gyrs, $Q_2=150$ and initial $e_4=0.0001$. On the top, evolution of the eccentricities and inclinations of the three moons in a case where the 2:1 mean motion resonance between Ariel and Umbriel is disrupted (case A). In the middle, evolution of the eccentricities and inclinations of the three moons in a case where also Titania is captured into mean motion resonance with Ariel and Umbriel (case B). On the bottom, evolution of the ratios of the mean motions in both cases.}
   \label{fig:321cases}
\end{figure*}

\subsection{Crossing of 3:2:1 resonance with Titania}
\label{subsec:321res}

For exploring the crossing of the 3:2:1 resonant chain, we used the same scheme already presented in Sect.~\ref{subsec:421res} for the 4:2:1 resonance. The only difference is that the initial value of the semi-major axis of Titania is set to $a_4=12.310~R_U$, in order to initially have $n_3/n_4< 1.5$, and $t_{\alpha,4}=5.7$ Gyrs.

As before, we divide the simulations of the various setups between those for which the 2:1 mean motion resonance between Ariel and Umbriel breaks down (case A) and those for which is preserved (case B). From \figurename~\ref{fig:escapes} (bottom panel), we can appreciate how the probability of escape from the 2:1 resonance is much larger than that we obtained with the previous resonant chain. This is likely due to the divergent nature of the resonant crossing. In fact, differently from two-body resonances, for three-body resonances it is possible to have a resonant capture with a divergent encounter, however the probability is generally smaller than that for a convergent encounter (for comparison, see \citealp{LARI-etal_2020,LARI-etal_2023}). More importantly, we can note that, differently from the 4:2:1 case, also for low values of $Q_2$ the disruption of the 2:1 resonance between Ariel and Umbriel is possible.

In \figurename~\ref{fig:321cases} (top panel), we report an example of this kind of evolution. For such simulations, there is not a chaotic variation of the orbital elements before the three-body resonance encounter. At the crossing of the resonance with Titania and the breaking of the 2:1 resonance, the orbital elements experience just a kick, which pushes both eccentricities and inclinations to new mean values. In the simulation presented in the figure, the variation of the orbital elements at the exit of the 2:1 resonance is quite mild, with only a small kick in the inclination of Ariel and the appearance of free eccentricities for Umbriel and Titania. This kind of evolution is very favorable, as moons' inclinations remain low and there is no need to invoke further dynamical mechanisms to damp them (see \citealp{CUK-etal_2020} and Sect.~\ref{sec:discu}). However, this particular outcome is not extremely likely, as the probability of breaking the 2:1 resonance for $Q_2\le 150$ is quite small (always less than $30\%$).

Unlike the 4:2:1 resonant crossing, simulations of case B can result either in the formation of a three-body resonant chain or preservation of the 2:1 resonance between Ariel and Umbriel without the capture of Titania. In \figurename~\ref{fig:321cases} (middle panel), we report a simulation of the first kind. The behavior is similar to the case B for the 4:2:1 resonance, apart from the forced values of the eccentricities of the three moons, which differ because of the resonances involved (3:2 between Umbriel and Titania, instead of 2:1). As before, this kind of evolution is not compatible with the current orbital configuration of the system.

We took a subset of the exit values of the moons' orbital elements from simulations of case A and we mapped them in \figurename~\ref{fig:incecc} (bottom panel). The number of points is larger than that for the 4:2:1 thanks to the higher escape probability ($>60\%$ already at $Q_2=200$), while the overall distribution of both eccentricities and inclinations is quite similar. In fact, final values of the orbital elements are spread stochastically by the crossing of the resonance jungle. Only the eccentricity of Ariel is flattened toward its current low value, because it is quickly damped by its assumed strong tidal friction.

In the end, the crossing of the 3:2:1 resonant chain is more efficient to break the 2:1 resonance and make it possible for the system to reach its current non-resonant configuration. Although not very likely, some simulations show evolutions that do not excite too much eccentricities and inclinations. However, after this resonance crossing, the moons migrate over quite a large range before reaching their current state (see \figurename~\ref{fig:AUTescape}, right plot); during this evolution, we expect them to cross other weaker resonances (not studied here) and undergo tidal damping.

\section{Discussion}
\label{sec:discu}

In this work, we showed that the crossing of a three-body resonance with Titania could have disrupted the 2:1 mean motion resonance between Ariel and Umbriel, which was previously thought to be unbreakable \citep{TITTEMORE-WISDOM_1990}. We found that the probability of breaking down the 2:1 resonance is higher for the 3:2:1 resonant crossing than the 4:2:1. Moreover, the probability increases significantly when taking higher values for the quality factor of Ariel. However, high values of $Q_2$ impact negatively the final moons' orbital elements. For $Q_2 \gtrsim 200$ (fixing $k_{2,2}=0.15$), Ariel and Umbriel trigger a chaotic variation of their eccentricities and inclinations, which end up to values generally higher than their current ones.

In this section, we address this and other critical issues of the Uranian moons' evolution we proposed.

\subsection{Orbital evolution after the breaking of the 2:1 resonance}
\label{subsec:orbev}

Since, after the breaking of the 2:1 resonance, the inclinations and eccentricities of the moons could have reached values much higher than their current ones, it is necessary that in the following evolution they damped to smaller values. A similar orbital excitation for all moons was obtained by \citet{CUK-etal_2020} when they investigated the encounter with the 5:3 mean motion resonance between Ariel and Umbriel. The authors proposed different mechanisms to damp the values of moons' eccentricities and inclinations after the escape from the 5:3 resonance.

Considering all case A simulations with $Q_2\le 250$ and initial $e_4=0.0001$ for the 3:2:1 resonant crossing, we obtain the following average values for the orbital elements of Ariel, Umbriel and Titania at the exit from the 2:1 resonance: $e_2=0.0015$, $e_3=0.0148$, $e_4=0.0099$, $I_2=0.850^\circ$, $I_3=0.795^\circ$ and $I_4=0.378^\circ$ (see \figurename~\ref{fig:incecc}, bottom panel). We observe that the eccentricity of Umbriel is generally higher than the others. This is probably due to the combined first-order resonant terms with both Ariel and Titania, which make the forcing effect greater on the satellite in the middle, such as in the case of the Laplace resonance between the Galilean satellites (see, e.g., \citealp{Lari_2018}). This could explain why today's value of Umbriel's eccentricity is higher than those of the other moons (see \tablename~\ref{tab:param}).

After the breaking of the 2:1 mean motion resonance, moons' free eccentricities can be efficiently damped by tidal dissipation within the satellites (see Eq.~\ref{eqn:tides}). However, this could require dissipation parameters $k_{2,i}/Q_i$ larger than the one used in our simulations (apart from Ariel). If we assume that the system has $400$ Myrs from the breaking of the 2:1 resonance to the current epoch, then we can compute the value of the dissipative parameters $k_{2,i}/Q_i$ necessary to damp the eccentricities to their current values. For Umbriel, we obtain $k_{2,3}/Q_3\approx 1.5\times 10^{-4}$, which is $5$ times larger than the value we used in our simulations. For Titania, we obtain $k_{2,4}/Q_4 \approx 3\times 10^{-3}$ in the scenario of the crossing of the 4:2:1 resonant chain and $k_{2,4}/Q_4 \approx 1.5\times 10^{-3}$ for the 3:2:1 resonant chain, which are $100$ and $50$ times larger than the value we used in our simulations, respectively.

A high value of the dissipative parameter of Titania was obtained also by \citep{CUK-etal_2020}, which proposed the presence of a subsurface ocean on Titania in order to support their findings (see also \citealp{CASTILLO-etal_2023}). Nevertheless, as shown by the same authors, secular resonances can redistribute the values of the eccentricities between the satellites, so that a higher eccentricity can be in part reduced by the passage through this kind of resonances. Therefore, it is possible that the values of $k_{2,i}/Q_i$ required to damp the eccentricities could be smaller than those computed above.

Unlike eccentricities, inclinations are more difficult to be damped, as tides are generally less efficient. However, if moons' obliquities are (ore were) significantly different from zero, obliquity tides (not included in our model) could damp inclinations more efficiently, especially in presence of subsurface oceans \citep{DOWNEY-etal_2020}. Furthermore, \citet{CUK-etal_2020} showed that the high inclinations they obtained at the exit of the 5:3 resonance could have been reduced through the combined action of secular and spin-orbit resonances.

Finally, in the evolution that we proposed, other resonances can play a role after the breaking of the 2:1 mean motion resonance between Ariel and Umbriel. From \figurename~\ref{fig:AUTescape}, we observe that after the 4:2:1 resonant crossing, the three moons do not encounter any major resonance, apart from the 5:3 between Ariel and Umbriel. Instead, after the 3:2:1 resonant crossing, there is also a three-body resonance encounter and a divergent 2:1 resonance encounter between Umbriel and Titania. While we do not expect any capture into these two resonances, they could contribute to further orbital excitation or damping of the satellites.

Values of eccentricities and inclinations of Ariel and Umbriel higher than their current ones are then required to pass the following 5:3 mean motion resonance between Ariel and Umbriel without being captured \citep{GOMES-CORREIA_2023,GOMES-CORREIA_2024b}. Moreover, the crossing of this resonance could have resulted in a final kick that could have reduced part of the pre-resonant values of the eccentricities and inclinations \citep{GOMES-CORREIA_2024b}.

\subsection{Tidal heating within the moons}
\label{subsec:therm}

The capture into the 2:1 mean motion resonance and consequent increase of moons' eccentricities enhanced the tidal heating within the satellites. The energy dissipated can be calculated through the following formula (e.g., \citealp{PEALE-etal_1979}):
\begin{equation}
    \label{eq:endis}
    \dot{E}_i=\frac{21}{2}\frac{k_{2,i}}{Q_i}\frac{n_i^5R_i^5}{\mathcal{G}}e_i^2\;.
\end{equation}

As described in Sect.~\ref{sec:2bres}, during the lock into resonance, the tidal heating within Ariel is large enough to account for its resurfacing. This outcome agrees with the results presented by \citet{NIMMO_2023}. Following the scenario proposed by the author, we set $t_{\alpha,2}=2.3$ Gyrs in order to get $Q_{U,2}\approx 1\,000$ at the time of the breaking of the 2:1 resonance. Inserting the value of the equilibrium eccentricity of Ariel (see Eq.~\ref{eq:e2eq}) into Eq.~\eqref{eq:endis} yields a total tidal heating of about $250$ GW, which corresponds to a past heat flux of $60$ mW/m$^2$, in line with the predictions for this moon \citep{PETERSON-etal_2015,NIMMO_2023}. Values of $t_{\alpha,2}$ between $2.3$ and $6.0$ Gyrs still provide an admissible amount of tidal heating (between $100$ and $250$ GW). Simulations in this range agree qualitatively with the ones obtained with $t_{\alpha,2}=2.3$ Gyrs and presented in Sect.~\ref{sec:break}, as they differ just for a small rescaling of $Q_{U,2}$. However, the reported analysis on the values of $Q_2$ should be rescaled by the same factor of $Q_{U,2}$, as most dynamical resonant features depend on the ratio between the two quality factors (see, e.g., Eqs.~\ref{eqn:ciDi} and~\ref{eq:e2eq}).

Unlike Ariel, Umbriel's surface does not show any evidence of past geological activity. This is not in contradiction with the capture and evolution into the 2:1 mean motion resonance. In fact, during the regular evolution within the resonance, the tidal heating within Umbriel is much lower than the one computed for Ariel. Considering that the two moons have similar masses and sizes, that during the resonance $n_2/n_3=2$ and that the equilibrium eccentricity of Umbriel is about $4$ times smaller than the one of Ariel (see \figurename~\ref{fig:21capture}), we obtain that the energy dissipation within Umbriel was at least $500$ times smaller than that of Ariel.

For Titania, we observe a certain excitation of its eccentricity at the crossing of the three-body resonance. From the average value of $e_4$ at the exit of the 2:1 resonance, we obtain a maximum tidal heating of $2$ GW (considering $a_4=14~R_U$ at the breaking of the 2:1 resonance through the 3:2:1 crossing and a high $k_{2,4}/Q_4=10^{-3}$). Analyzing tectonic features on the surface of Titania, \citet{BEDDINGFIELD-etal_2023} predicted past heat fluxes on the moon to be between $5$ and $12$ mW/m$^2$, which would require tens of GW of energy dissipation. Therefore, the crossing of the 4:2:1 or 3:2:1 resonant chains between Ariel, Umbriel and Titania cannot be responsible of the resurfacing of Titania. In fact, there are not clear estimations of the age of the geological features on Titania, but they are probably older than the last resurfacing episode of Ariel and the formation of the youngest coronae on Miranda \citep{ZAHNLE-etal_2003}.

\subsection{Consequences on Miranda}
\label{subsec:miran}

In the evolution presented in Sect.~\ref{sec:break}, we focused on Ariel, Umbriel and Titania, as they are the three moons involved in the three-body resonant chain. However, the passage through such a configuration has an impact on the whole system. In particular, being the smallest of the five regular satellites of Uranus, Miranda could have undergone a large orbital excitation. This indirect outcome was already observed by \citet{CUK-etal_2020} during the crossing of the 5:3 resonance, which they found could be responsible for both the current high inclination and the formation of the young coronae observed on the surface of Miranda.

\begin{figure}
\centering
\includegraphics[scale=0.48]{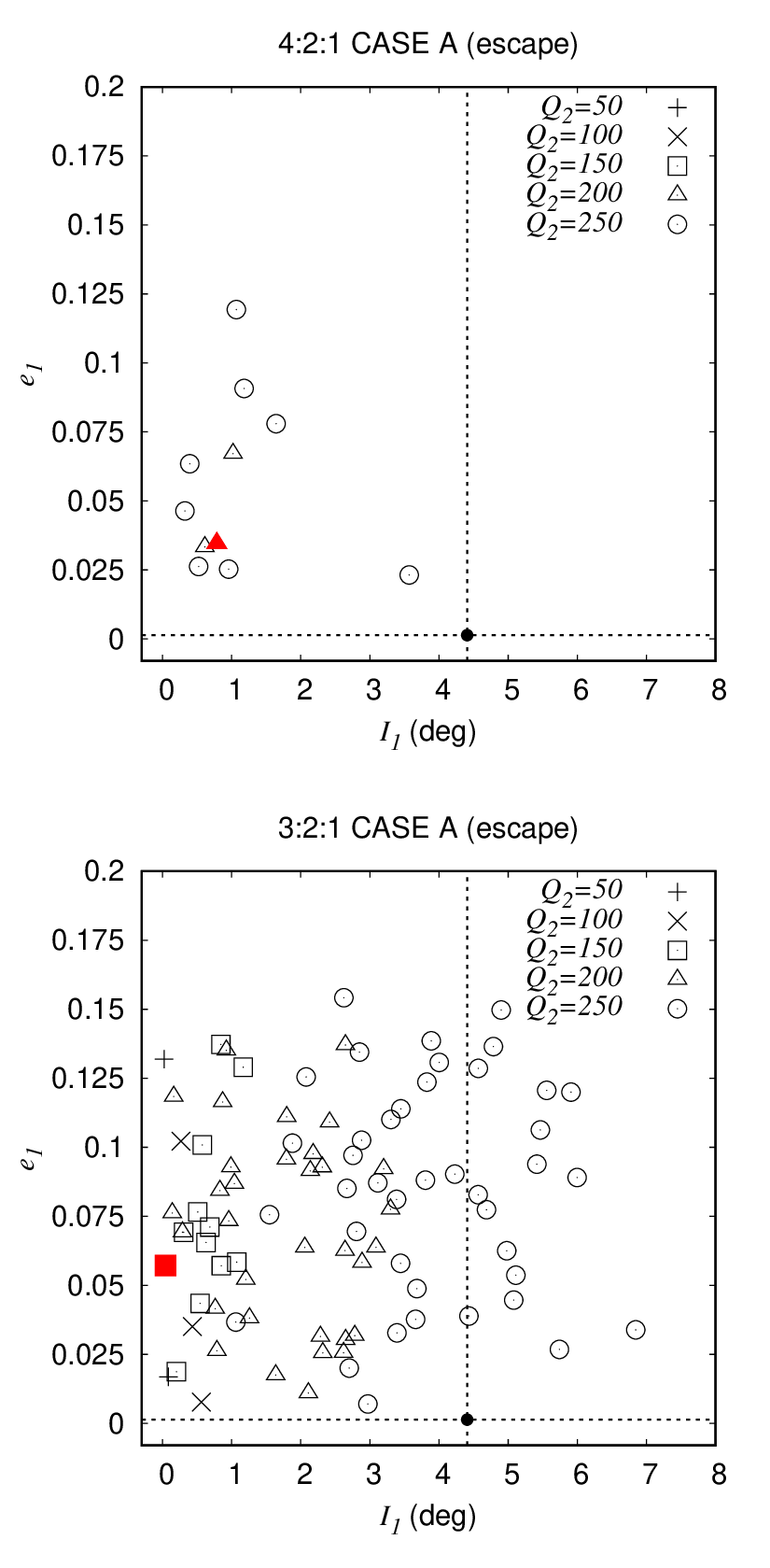}
\caption{Mean orbital eccentricities and inclinations of Miranda obtained from the numerical simulations (only the ones with initial $e_4=0.0001$ and $Q_2\leq 250$) after the disruption of the 2:1 mean motion resonance, both in the case of 4:2:1 resonant crossing (top) and 3:2:1 resonant crossing (bottom). The values are taken when the system reached $n_2/n_3=1.94$ (see text in Sect.~\ref{subsec:421res}). The dashed lines indicate the current values of the mean orbital elements. The red points show the orbital elements of the simulations presented in \figurename~\ref{fig:421cases} and~\ref{fig:321cases}.}
\label{fig:inceccM}
\end{figure}

As done for Ariel, Umbriel and Titania, we looked at the eccentricity and inclination of Miranda at the breaking of the 2:1 resonance. Their values are reported in \figurename~\ref{fig:inceccM} for both the crossing of the 4:2:1 and 3:2:1 resonant chains.

We obtain that the final eccentricity of Miranda can become as high as $0.1542$, with an average value of $0.0773$ (in the case of 3:2:1 resonant crossing). In order to damp this mean value in $400$ Myrs it is necessary that $k_{2,1}/Q_1\approx 2\times 10^{-5}$, which is a quite low value (although slightly larger than what we assumed in our simulations). Analyzing the coronae on the surface of Miranda, \citet{BEDDINGFIELD-etal_2022} predicted past heat fluxes to be between $35$ and $140$ mW/m$^2$, corresponding to global tidal heating of tens of GW. Considering $e_1=0.0773$, this would require $k_{2,1}/Q_1 > 10^{-4}$, which is reasonable if melting within the moon occurred during the extreme excitation of its orbit (see \citealp{CUK-etal_2020,CASTILLO-etal_2023}).

\begin{figure}
\centering
\includegraphics[scale=0.6]{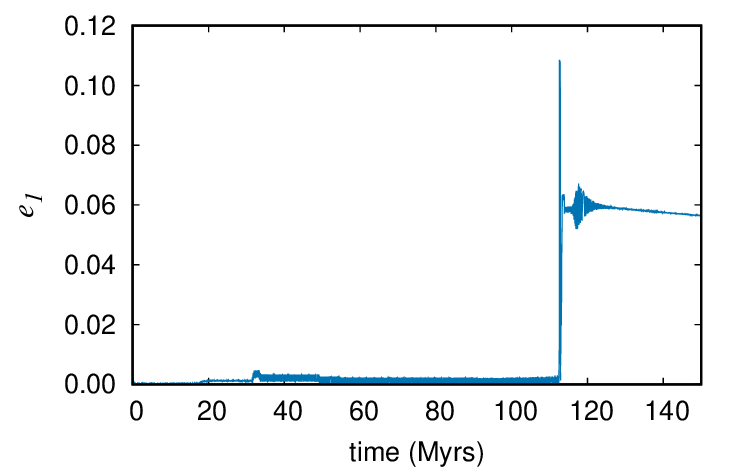}
\caption{Evolution of the eccentricity of Miranda in a case where the 2:1 mean motion resonance between Ariel and Umbriel is disrupted through the 3:2:1 resonant crossing.}
\label{fig:eccmir}
\end{figure}

While the resurfacing of Ariel could have occurred during the whole timespan of the lock into mean motion resonance with Umbriel, this large tidal heating event on Miranda would have happened shortly thereafter, at (or just after) the breaking of the 2:1 resonance (see \figurename~\ref{fig:eccmir}), i.e., exactly between $200$ and $600$ Myrs ago. This time interval matches the estimation of the age of the youngest coronae on Miranda \citep{BEDDINGFIELD-etal_2022}.

Finally, the inclination of Miranda can reach few degrees after the breaking of the 2:1 resonance, with an average exit value of $2.505^\circ$ (considering simulations with $Q_2\le 250$ in the case of 3:2:1 resonant crossing). Therefore, the scenario we presented can explain the current high inclination of the moon. From \figurename~\ref{fig:inceccM}, we can note that the final inclination results in a systematically larger value when taking higher values of $Q_2$. This is due to the larger increase of Ariel's inclination during the lock into resonance with Umbriel which impacts greatly on the inclination of Miranda (see Sect.~\ref{sec:2bres}).

\subsection{Other scenarios for a fast-migration regime}
\label{subsec:other}

Our exploration of the orbital history of the Uranian satellites assumed a fast orbital migration of Ariel and Titania, while we considered low migration rates (or equivalently high $Q_{U,i}$) for the other moons.

Capture into the 2:1 mean motion resonance between Ariel and Umbriel requires that Ariel migrates faster than the outer moon Umbriel. This is a natural outcome if the value of $Q_{U,i}$ is equal for $i=2,3$ (e.g., if the quality factor of Uranus is constant), since the two satellites have comparable masses. However, from tidal resonance locking theory (see Sect.~\ref{subsec:resloc}), it is possible that Umbriel is locked with an interior mode of Uranus that evolves faster than the one associated with Ariel \citep{FULLER-etal_2016}, or even with the same timescale \citep{LAINEY-etal_2020}. In these cases, Ariel and Umbriel would have had only divergent resonant encounters or no resonant encounters at all, which makes it difficult to explain resurfacing of Ariel through tidal heating. Therefore, the 2:1 mean motion resonance between the two moons appears to be unavoidable when considering the thermal-orbital evolution of the satellite system in the context of a low value of $Q_U$ \citep{NIMMO_2023,JACOBSON-PARK_2025}.

We did not consider fast migration for Miranda and Oberon, as these two moons do not enter directly into the resonant orbital evolution we proposed. However, assessing their migration rates is essential for the reconstruction of the entire thermal-orbital history of the Uranian system \citep {NIMMO_2023}. In particular, knowing the orbital expansion rate of Miranda is necessary to correctly investigate and place the epoch of the crossing of the 3:1 mean motion resonance with Umbriel \citep{TITTEMORE-WISDOM_1989}, which we did not consider in this study.

Finally, when assuming a low value of the quality factor of Uranus, we had to consider a large variation of $Q_U$ during the whole history of the moon system (see Eq.~\ref{eq:Qreslock}). In fact, $Q_U\approx 1\,000$ cannot be sustained for billions of years if we assume that the satellites formed from a planetary disk generated by an early giant impact \citep{IDA-etal_2020,MORBIDELLI-etal_2012,WOO-etal_2022}. However, if the moons formed much later, for example due to the disruption of an ancient large satellite (\citealp{BOUE-LASKAR_2010,SAILLENFEST-etal_2022}; see \citealp{STROM_1987} and \citealp{CASTILLO-etal_2023} for moons' age estimates based on crater counting), then $Q_U$ could have been small for the whole moons' history.

\section{Conclusion}
\label{sec:concl}

A low value of the quality factor of Uranus revolutionizes the classical orbital history depicted for the Uranian moons (e.g., \citealp{DERMOTT-etal_1988}). In particular, the fast migration of some moons alters the timing of the resonant encounters and even brings into play orbital resonances previously discarded or never considered.

In this work, we analyzed the 2:1 mean motion resonance between Ariel and Umbriel, which was discarded by all past orbital reconstruction of the Uranus system \citep{TITTEMORE-WISDOM_1990}. In particular, assuming $Q_{U}\approx 1\,000$ \citep{NIMMO_2023}, it follows that the resonance must have been crossed in the past and even quite recently (less than one billion years ago).

As the capture into the 2:1 resonance is certain if the pre-resonant encounter eccentricities of Ariel and Umbriel were small (see also \citealp{TITTEMORE-WISDOM_1990,MURRAY-DERMOTT_2000}), we investigated a possible dynamical mechanism to break it down, so that the system could have eventually reached its current non-resonant configuration. More precisely, we proposed a further resonant encounter with Titania, so that the system crossed a three-body resonance between the three moons. Assuming a quality factor of Uranus at the frequency of Ariel $Q_{U,2}$ of order $1\,000$, it is required that the same quality factor at the frequency of Titania must be one or two orders of magnitude smaller.

The probability that the three-body resonance crossing succeeds in breaking the 2:1 resonance depends on the dynamical configuration of the system and the values of some parameters. In our analysis, we considered the crossing of the 3:2:1 and 4:2:1 resonant chains between Ariel, Umbriel, and Titania. We investigated the impact of the tidal quality factor $Q_2$ of Ariel (for a fixed relatively high value of $k_{2,2}$) and the initial eccentricity of Titania. We found that higher values of $Q_2$ increase the probability of disruption of the 2:1 resonance. However, high values of $Q_2$ also cause Ariel and Umbriel to trigger inclination resonances, and induce chaotic variations of their orbital elements due to resonance overlapping. Therefore, we focused on the best trade-off between low excitation of eccentricities and inclinations of the moons and high probability of escape from the resonance.

Due to its divergent nature, the scenario involving the 3:2:1 resonance is more favorable, as for $Q_2\ge 200$ the exit probability reaches values greater than $60\%$ (see \figurename~\ref{fig:escapes}). While a few simulations result in small final eccentricities and inclinations, the general behavior is an excitation of both these orbital elements, whose final values span quite a wide range, as shown in \figurename~\ref{fig:incecc}. The subsequent evolution could lead to a significant reduction of these values, even though the damping of the inclinations might be more difficult \citep{CUK-etal_2020}. Nevertheless, the relatively high eccentricity and inclination of Umbriel that we obtained could explain why this moon has the largest eccentricity of the system and how Ariel and Umbriel managed to cross the 5:3 resonance without being captured \citep{GOMES-CORREIA_2023,GOMES-CORREIA_2024a}.

The capture of Ariel and Umbriel into the 2:1 resonance has strong implications on the thermal history of the Uranian moons. More precisely, the resonance forces Ariel's eccentricity to values that cause strong tidal heating, matching geophysical estimates of the past heat flux of this moon \citep{PETERSON-etal_2015,NIMMO_2023}. While the determination of the epoch of entrance into resonance depends on the initial configuration of the moons, the time of exit can be obtained from the assumed value of $Q_{U,2}$. Considering the range derived by \citet{NIMMO_2023} and the quality factor variation expected by tidal resonance locking \citep{FULLER-etal_2016}, we estimated that the exit from the resonance should have occurred between $200$ and $600$ million years ago. Therefore, the resurfacing of Ariel must be dated before this dynamical event, possibly between a few billion (or even less) and a few hundred million years ago. This range agrees with the estimated age of the youngest geological features of Ariel \citep{PETERSON-etal_2015,SCHENK-MOORE_2020}. Furthermore, since the three-body resonant crossing produces a large increase in the eccentricity of Miranda, we obtain a phase of extreme tidal heating within the little moon exactly between $200$ and $600$ million years ago. This range closely matches the predicted age of the formation of the youngest coronae on Miranda \citep{BEDDINGFIELD-etal_2022}.

Finally, the fast migration of the moons changes the timing of the mean motion resonances previously analyzed \citep{TITTEMORE-WISDOM_1988,TITTEMORE-WISDOM_1989,CUK-etal_2020,GOMES-CORREIA_2024b}. In particular, considering the same $t_{\alpha,2}$ we derived in this study, the crossing of the 5:3 mean motion resonance between Ariel and Umbriel should have occurred between $13$ and $35$ million years ago. The proximity of this resonance to the current epoch makes it necessary to revise previous results in the literature. In particular, past studies relied on hundreds of millions of years to damp the orbital elements excited by the 5:3 resonant crossing. If the same must be accomplished in a few tens of millions of years, then the dissipative parameters of the moons must be $10$ or even $100$ times smaller than the values previously considered. This has direct repercussions on our knowledge of the moons' interior and the possibility that they currently harbor subsurface oceans \citep{CASTILLO-etal_2023}.

The future measurements of the Uranus Orbiter and Probe mission will hopefully provide precise estimations of the dissipative parameters of the moons and of Uranus at the different orbital frequencies. Their values will be essential to constrain the orbital and thermal history of the Uranian satellites. In particular, the measure of $Q_{U,2}$ will ultimately allow verification of the high tidal dissipation scenario proposed by \citet{NIMMO_2023} and the consequent crossing of the 2:1 resonance, while the measure of $Q_{U,4}$ will provide evidence or confutation of the dynamical mechanism we proposed in this work to disrupt the resonance.

\begin{acknowledgements}
This research has been developed under the ASI/UniBo-CIRI agreement n. 2024-5-HH.0. M.R. is grateful to the Gruppo Nazionale per la Fisica Matematica -- Istituto Nazionale di Alta Matematica (GNFM-INdAM).
\end{acknowledgements}

\bibliographystyle{aa} 
\bibliography{refs}

\appendix

\section{Averaged dynamical model}
\label{sec:appA}

This section is devoted to outlining the main steps leading to the Hamiltonian in Eqs.~\eqref{eqn:ham}--\eqref{eqn:hamD}, as obtained using a computer algebra system.

\subsection{Orbital element expansion of the Hamiltonian expressed in Cartesian coordinates}
\label{subsec:expham}

We start from the Hamiltonian \eqref{eqn:ham} of the 6-body planetary problem expressed in Cartesian coordinates, whose components are given by
\begin{equation}
    \label{eqn:hamCart}
    \begin{aligned}
    \mathcal{H}_K&=\sum_{i=1}^5\left(\frac12\frac{\norm{\boldsymbol{p}_i}^2}{\beta_i}-\frac{\mu_i\beta_i}{\norm{\boldsymbol{r}_i}}\right)\;,\\
    \mathcal{H}_O&=\sum_{i=1}^5\frac{\mathcal{G}m_Um_i}{\norm{\boldsymbol{r}_i}}J_2\left(\frac{R_U}{\norm{\boldsymbol{r}_i}}\right)^2\mathcal{P}_2\left(\frac{\boldsymbol{r}_i}{\norm{\boldsymbol{r}_i}}\cdot\boldsymbol{s}_U\right)\;,\\
    \mathcal{H}_I&=\sum_{1\le i< j\le 5}\frac{\boldsymbol{p}_i\cdot\boldsymbol{p}_j}{m_U}\;,\\
    \mathcal{H}_D&=-\sum_{1\le i< j\le 5}\frac{\mathcal{G}m_im_j}{\norm{\boldsymbol{r}_i-\boldsymbol{r}_j}}\;,
    \end{aligned}
\end{equation} 
where $(\boldsymbol{r}_i,\boldsymbol{p}_i)$ are the Uranus-centric canonical variables of the satellites \citep{LASKAR-ROBUTEL_1995}, $\mathcal{P}_2(x)=(3x^2-1)/2$ is the Legendre polynomial of degree two, and $\boldsymbol{s}_U$ is the unit spin vector of Uranus. The five integrable two-body problems in $\mathcal{H}_K$ can be immediately written in terms of their Keplerian energy, that is Eq.~  \eqref{eqn:hamkep}. We pass to orbital elements in $\mathcal{H}_I$ via
\begin{equation}
    \label{eqn:pielem}
     \boldsymbol{p}_i=\beta_i\mathcal{R}_i\left(-\frac{n_ia_i\sin f_i}{\sqrt{1-e_i^2}},\frac{n_ia_i(\cos f_i+e_i)}{\sqrt{1-e_i^2}},0\right)^T\;,
\end{equation}
where $f_i$ is the $i$-th true anomaly, related to the mean anomaly $\ell_i$ through Bessel functions of the first kind $\mathcal{J}_{\nu}$ as
\begin{equation}
    \label{eqn:bessel}
    \begin{aligned}
    \cos f_i&= \frac{2(1-e_i^2)}{e_i}\sum_{\nu=1}^{\infty}\mathcal{J}_{\nu}(\nu e_i)\cos(\nu\ell_i)-e_i\;,\\
    \sin f_i&=\sqrt{1-e_i^2}\sum_{\nu=1}^{\infty}\frac12\left(\mathcal{J}_{\nu-1}(\nu e_i)-\mathcal{J}_{\nu+1}(\nu e_i)\right)\sin(\nu\ell_i)\;.
    \end{aligned}
\end{equation}
The matrix
\begin{equation}
    \label{eqn:rotmat}
    \mathcal{R}_i=\begin{pmatrix}
     \mathrm{c}\Omega_i\,\mathrm{c}\omega_i - \mathrm{s}\Omega_i\,\mathrm{c}I_i\,\mathrm{s}\omega_i & -\mathrm{c}\Omega_i\,\mathrm{s}\omega_i-\mathrm{s}\Omega_i\,\mathrm{c}I_i\,\mathrm{c}\omega_i & \mathrm{s}\Omega_i\,\mathrm{s}I_i \\
     \mathrm{s}\Omega_i\,\mathrm{c}\omega_i + \mathrm{c}\Omega_i\,\mathrm{c}I_i\,\mathrm{s}\omega_i & -\mathrm{s}\Omega_i\,\mathrm{s}\omega_i+\mathrm{c}\Omega_i\,\mathrm{c}I_i\,\mathrm{c}\omega_i & -\mathrm{c}\Omega_i\,\mathrm{s}I_i\\
     \mathrm{s}I_i\,\mathrm{s}\omega_i & \mathrm{s}I_i\,\mathrm{c}\omega_i & \mathrm{c}I_i
    \end{pmatrix}
\end{equation}
is the standard rotation matrix from perifocal to equatorial coordinates, in which $\mathrm{c}\cdot$ and $\mathrm{s}\cdot$ are abbreviations for cosine and sine, respectively, and $\omega_i=\varpi_i-\Omega_i$ is the argument of pericenter of the satellite $i$. Eq.~\eqref{eqn:pielem} is only valid in a very specific reference frame, that is, the orbital frame, which is the inertial one used in the present work. The same equation allows us to introduce the modified Delaunay variables defined in Eq.~\eqref{eqn:modDel}. Regarding $\mathcal{H}_O$, we use 
\begin{equation}
    \label{eqn:dotprod}
    \boldsymbol{r}_i\cdot\boldsymbol{s}_U=\norm{\boldsymbol{r}_i}\sin I_i\sin(f_i+\omega_i)
\end{equation}
and
\begin{equation}
    \label{eqn:rielem}
    \norm{\boldsymbol{r}_i}=\frac{a_i(1-e_i^2)}{1+e_i\cos f_i}\;,
\end{equation}
along with Eq.~\eqref{eqn:bessel}. Finally, $\mathcal{H}_D$ is based on Ellis-Kaula's expansion of the direct part of the disturbing function in the three-body problem, as described in \citet{MURRAY-DERMOTT_2000} (see also references therein), adapted to the Hamiltonian planetary case. For a given specific order in the eccentricities and inclinations, the expansion is a finite series composed of trigonometric polynomials of the form $A(a_i,a_j,e_i,e_j,s_i,s_j)\cos\phi$, for each valid argument
\begin{equation}
    \label{eqn:arg}
    \phi=j_1\lambda_i+j_2\lambda_j+j_3\varpi_i+j_4\varpi_j+j_5\Omega_i+j_6\Omega_j\;,\quad j_l\in\mathbb{Z}\;,
\end{equation}
i.e., that satisfies D'Alembert rules. Therefore, it is possible to list all such arguments $\phi$ (cf. \citealp{MURRAY-DERMOTT_2000}, Appendix B) and compute the corresponding coefficients $A$ using Ellis-Murray's formulas (\citealp{MURRAY-DERMOTT_2000}, Section 6.6). 

Upon truncating the expansions to order $e_i^2$ and $s_i^2$, we choose values of $p_1$, $p_2$ and $q_1$, $q_2$ corresponding to the 4:2:1 and 3:2:1 Ariel-Umbriel-Titania mean motion resonances. By construction, the approach is completely general and applies to any other commensurability of order $q_1,q_2\le2$. Ultimately, the averaging
\begin{equation}
    \label{eqn:avg}
    \langle\,\cdot\,\rangle_{p_1,p_2,q_1,q_2}=\frac{1}{(2\pi)^5}\int_{[0,2\pi]^5}\cdot\;\mathrm{d}\lambda_1\ldots\mathrm{d}\lambda_5\;,
\end{equation}
applied to $\mathcal{H}_O$, $\mathcal{H}_I$, and $\mathcal{H}_D$ in Eq.~\eqref{eqn:hamCart} yields Eqs.~\eqref{eqn:hamobl}--\eqref{eqn:hamD}, where $\langle\,\cdot\,\rangle_{p_1,p_2,q_1,q_2}$ denotes the exclusion from the averaging operation of the corresponding resonant combinations, as discussed in Sect.~\ref{subsec:avgham}.

\subsection{Terms of the direct part}
\label{subsec:HD}

Here, we report the secular and resonant terms that constitute Eq.~\eqref{eqn:hamD}, obtained as a result of the algebraic manipulations illustrated in Sect.~\ref{subsec:expham}.
\begin{equation}
\label{eqn:hamDsec}
\begin{aligned}
    \mathcal{H}_{D,\text{sec}}=&-\sum_{1\le i<j\le 5}\frac{\mathcal{G}m_im_j}{a_j}\big(F_1+F_2(e_i^2+e_j^2)+F_3(s_i^2+s_j^2)\\
    &+F_4e_ie_j\cos(\varpi_i-\varpi_j)+F_5s_is_j\cos(\Omega_i-\Omega_j)\big)\;,
    \end{aligned}
\end{equation}
\begin{equation}
\label{eqn:hamD21}
\begin{aligned}
    \mathcal{H}_{D,\text{2:1},ij}=&-\frac{\mathcal{G}m_im_j}{a_j}\big(F_6 e_i\cos(2\lambda_j-\lambda_i-\varpi_i)\\
    &+F_7e_j\cos(2\lambda_j-\lambda_i-\varpi_j)\\
    &+F_8e_i^2\cos(4\lambda_j-2\lambda_i-2\varpi_i)\\
    &+F_9e_j^2\cos(4\lambda_j-2\lambda_i-2\varpi_j)\\
    &+F_{10}e_ie_j\cos(4\lambda_j-2\lambda_i-\varpi_i-\varpi_j)\\
    &+F_{11}s_i^2\cos(4\lambda_j-2\lambda_i-2\Omega_i)\\
    &+F_{12}s_j^2\cos(4\lambda_j-2\lambda_i-2\Omega_j)\\
    &+F_{13}s_is_j\cos(4\lambda_j-2\lambda_i-\Omega_i-\Omega_j)\big)\;,\\
    &\hspace{-1.4cm} (i,j)=(2,3),(3,4)\;,
\end{aligned}
\end{equation}
\begin{equation}
\label{eqn:hamD32}
\begin{aligned}
    \mathcal{H}_{D,\text{3:2},34}=&-\frac{\mathcal{G}m_3m_4}{a_4}\big(F_{14}e_3\cos(3\lambda_4-2\lambda_3-\varpi_3)\\
    &+F_{15}e_4\cos(3\lambda_4-2\lambda_3-\varpi_4)\\
    &+F_{16}e_3^2\cos(6\lambda_4-4\lambda_3-2\varpi_3)\\
    &+F_{17}e_4^2\cos(6\lambda_4-4\lambda_3-2\varpi_4)\\
    &+F_{18}e_3e_4\cos(6\lambda_4-4\lambda_3-\varpi_3-\varpi_4)\\
    &+F_{19}s_3^2\cos(6\lambda_4-4\lambda_3-2\Omega_3)\\
    &+F_{20}s_4^2\cos(6\lambda_4-4\lambda_3-2\Omega_4)\\
    &+F_{21}s_3s_4\cos(6\lambda_4-4\lambda_3-\Omega_3-\Omega_4)\big)\;,
\end{aligned}
\end{equation}
\begin{equation}
\label{eqn:hamD31}
\begin{aligned}
    \mathcal{H}_{D,\text{3:1},24}=&-\frac{\mathcal{G}m_2m_4}{a_4}\big(F_{22}e_2^2\cos(3\lambda_4-\lambda_2-2\varpi_2)\\
    &+F_{23}e_4^2\cos(3\lambda_4-\lambda_2-2\varpi_4)\\
    &+F_{24}e_2e_4\cos(3\lambda_4-\lambda_2-\varpi_2-\varpi_4)\\
    &+F_{25}s_2^2\cos(3\lambda_4-\lambda_2-2\Omega_2)\\
    &+F_{26}s_4^2\cos(3\lambda_4-\lambda_2-2\Omega_4)\\
    &+F_{27}s_2s_4\cos(3\lambda_4-\lambda_2-\Omega_2-\Omega_4)\big)\;,
\end{aligned}
\end{equation}
where
\begin{equation}
    \label{eqn:Lapcoeffsec}
    \begin{aligned}
        F_1&=\frac12 b_{1/2}^{(0)}(\alpha_{ij})\;,\\
        F_2&=\frac{\alpha_{ij}}{4}\left(\frac{\partial b_{1/2}^{(0)}(\alpha_{ij})}{\partial\alpha_{ij}}+\frac{\alpha_{ij}}{2}\frac{\partial^2b_{1/2}^{(0)}(\alpha_{ij})}{\partial\alpha_{ij}^2}\right)\;,\\
        F_3&=-\frac{\alpha_{ij}}{2}b_{3/2}^{(1)}(\alpha_{ij})\;,\\
        F_4&=\frac12\left(b_{1/2}^{(1)}(\alpha_{ij})-\alpha_{ij}\frac{\partial b_{1/2}^{(1)}(\alpha_{ij})}{\partial\alpha_{ij}}-\frac{\alpha_{ij}^2}{2}\frac{\partial^2 b_{1/2}^{(1)}(\alpha_{ij})}{\partial\alpha_{ij}^2}\right)\;,\\
        F_5&=-2F_3\;,
    \end{aligned}
\end{equation}
\begin{equation}
    \label{eqn:Lapcoeff21}
    \begin{aligned}
        F_6&=-2b_{1/2}^{(2)}(\alpha_{ij})-\frac12\alpha_{ij}\frac{\partial b_{1/2}^{(2)}(\alpha_{ij})}{\partial\alpha_{ij}}\;,\\
        F_7&=\frac12\left(3b_{1/2}^{(1)}(\alpha_{ij})+\alpha_{ij}\frac{\partial b_{1/2}^{(1)}(\alpha_{ij})}{\partial\alpha_{ij}}\right)\;,\\
        F_8&=\frac12\left(11b_{1/2}^{(4)}(\alpha_{ij})+\frac{7}{2}\alpha_{ij}\frac{\partial b_{1/2}^{(4)}(\alpha_{ij})}{\partial\alpha_{ij}}+\frac14\alpha_{ij}^2\frac{\partial^2b_{1/2}^{(2)}(\alpha_{ij})}{\partial\alpha_{ij}^2}\right)\;,\\
        F_9&=\frac14\left(19b_{1/2}^{(2)}(\alpha_{ij})+7\alpha_{ij}\frac{\partial b_{1/2}^{(2)}(\alpha_{ij})}{\partial\alpha_{ij}}+\frac12\alpha_{ij}^2\frac{\partial^2 b_{1/2}^{(2)}(\alpha_{ij})}{\partial\alpha_{ij}^2}\right)\;,\\
        F_{10}&=-\frac12\left(21b_{1/2}^{(3)}(\alpha_{ij})+7\alpha_{ij}\frac{\partial b_{1/2}^{(3)}(\alpha_{ij})}{\partial\alpha_{ij}}+\frac12\alpha_{ij}^2\frac{\partial^2 b_{1/2}^{(3)}(\alpha_{ij})}{\partial\alpha_{ij}^2}\right)\;,\\
        F_{11}&=\frac12\alpha_{ij}b_{3/2}^{(3)}(\alpha_{ij})\;,\\
        F_{12}&=F_{11}\;,\\
        F_{13}&=-2F_{11}\;,
    \end{aligned}
\end{equation}
\begin{equation}
    \label{eqn:Lapcoeff32}  
    \begin{aligned}
        F_{14}&=-3b_{1/2}^{(3)}(\alpha_{34})-\frac12\alpha_{34}\frac{\partial b_{1/2}^{(3)}(\alpha_{34})}{\partial\alpha_{34}}\;,\\
        F_{15}&=\frac12\left(5b_{1/2}^{(2)}(\alpha_{34})+\alpha_{34}\frac{\partial b_{1/2}^{(2)}(\alpha_{34})}{\partial\alpha_{34}}\right)\;,\\
        F_{16}&=\frac14\left(57b_{1/2}^{(6)}(\alpha_{34})+11\alpha_{34}\frac{\partial b_{1/2}^{(6)}(\alpha_{34})}{\partial\alpha_{34}}+\frac12\alpha_{34}^2\frac{\partial^2b_{1/2}^{(6)}(\alpha_{34})}{\partial\alpha_{34}^2}\right)\;,\\
        F_{17}&=13b_{1/2}^{(4)}(\alpha_{34})+\frac{11}{4}\alpha_{34}\frac{\partial b_{1/2}^{(4)}(\alpha_{34})}{\partial\alpha_{34}}+\frac{1}{8}\alpha_{34}^2\frac{\partial^2 b_{1/2}^{(4)}(\alpha_{34})}{\partial\alpha_{34}^2}\;,\\
        F_{18}&=-\frac12\left(55b_{1/2}^{(5)}(\alpha_{34})+11\alpha_{34}\frac{\partial b_{1/2}^{(5)}(\alpha_{34})}{\partial\alpha_{34}}+\frac12\alpha_{34}^2\frac{\partial^2 b_{1/2}^{(5)}(\alpha_{34})}{\partial\alpha_{34}^2}\right)\;,\\
        F_{19}&=\frac12\alpha_{34}b_{3/2}^{(5)}(\alpha_{34})\;,\\
        F_{20}&=F_{19}\;,\\
        F_{21}&=-2F_{19}\;,
    \end{aligned}
\end{equation}
\begin{equation}
    \label{eqn:Lapcoeff31}
    \begin{aligned}
        F_{22}&=\frac14\left(\frac{21}{2}b_{1/2}^{(3)}(\alpha_{24})+5\alpha_{24}\frac{\partial b_{1/2}^{(3)}(\alpha_{24})}{\partial\alpha_{24}}+\frac12\alpha_{24}^2\frac{\partial^2 b_{1/2}^{(3)}(\alpha_{24})}{\partial\alpha_{24}^2}\right)\;,\\
        F_{23}&=\frac14\left(\frac{17}{2}b_{1/2}^{(1)}(\alpha_{24})+5\alpha_{24}\frac{\partial b_{1/2}^{(1)}(\alpha_{24})}{\partial\alpha_{24}}+\frac12\alpha_{24}^2\frac{\partial^2 b_{1/2}^{(1)}(\alpha_{24})}{\partial\alpha_{24}^2}\right)\;,\\
        F_{24}&=-5b_{1/2}^{(2)}(\alpha_{24})-\frac52\alpha_{24}\frac{\partial b_{1/2}^{(2)}(\alpha_{24})}{\partial\alpha_{24}}-\frac14\alpha_{24}^2\frac{\partial^2 b_{1/2}^{(2)}(\alpha_{24})}{\partial\alpha_{24}^2}\;,\\
        F_{25}&=\frac12\alpha_{24}b_{3/2}^{(2)}(\alpha_{24})\;,\\
        F_{26}&=F_{25}\;,\\
        F_{27}&=-2F_{25}
    \end{aligned}
\end{equation}
depend on the Laplace coefficients
\begin{equation}
    \label{eqn:Lapcoefdef}
    b_s^{(k)}(\alpha_{ij})=\frac{1}{\pi}\int_{0}^{2\pi}\frac{\cos(k\psi)}{(1-2\alpha_{ij}\cos\psi+\alpha_{ij}^2)^s}\mathrm{d}\psi\;,\quad\alpha_{ij}=\frac{a_i}{a_j}<1\;,
\end{equation}
with $k=0,1,2,\ldots$ being an integer and $s=1/2,3/2,5/2,\ldots$ being a half-integer. We compute these coefficients and their derivatives following App.~\ref{subsec:laplcoeff}.

\subsection{Laplace coefficients}
\label{subsec:laplcoeff}

During the numerical simulations, the quantities \eqref{eqn:Lapcoefdef} are typically either kept constant, since they depend on the slowly time-varying ratio $\alpha_{ij}$, or computed numerically at each time step, along with their derivatives. To significantly improve computational speed and numerical precision, especially for $\alpha_{ij}\to 1$, other approaches have been proposed, such as the use of Chebyshev interpolations, successfully applied in \citet{LARI-etal_2020}. In our work, we achieve a good compromise between implementation simplicity and numerical accuracy by exploiting the series representation of Eq.~\eqref{eqn:Lapcoefdef} in powers of $\alpha_{ij}$ \citep{MURRAY-DERMOTT_2000}:
\begin{equation}
    \label{eqn:lapcoefseries}
    b_s^{(k)}(\alpha_{ij})=\sum_{l=0}^{\infty}B_l(s,k)\alpha_{ij}^{k+2l}\;,
\end{equation}
where $B_l(s,k)$ are coefficients that depend on the hypergeometric function. When the summation is truncated, Eq.~\eqref{eqn:lapcoefseries} reduces to an efficient polynomial evaluation. We safely truncate at order $l=100$ to ensure convergence at machine-precision level, as $\alpha_{ij}\lesssim0.8$ in all cases considered.

For completeness, the derivatives in Eqs.~\eqref{eqn:Lapcoeffsec}--\eqref{eqn:Lapcoeff31} are calculated using the recurrence relations \citep{BROUWER-CLEMENCE_1961}
\begin{equation}
\label{eqn:recurs}
\begin{aligned}
\frac{\partial b_s^{(k)}(\alpha_{ij})}{\partial\alpha_{ij}}&=s\left(b_{s+1}^{(k-1)}(\alpha_{ij})-2\alpha_{ij}b_{s+1}^{(k)}(\alpha_{ij})+b_{s+1}^{(k+1)}(\alpha_{ij})\right)\;,\\
\frac{\partial^l b_s^{(k)}(\alpha_{ij})}{\partial\alpha_{ij}^l}&=s\Bigg(\frac{\partial^{l-1}b_{s+1}^{(k-1)}(\alpha_{ij})}{\partial\alpha_{ij}^{l-1}}-2\alpha_{ij}\frac{\partial^{l-1}b_{s+1}^{(k)}(\alpha_{ij})}{\partial\alpha_{ij}^{l-1}}\\
&\;\;\;+\frac{\partial^{l-1}b_{s+1}^{(k+1)}(\alpha_{ij})}{\partial\alpha_{ij}^{l-1}}-2(l-1)\frac{\partial^{l-2}b_{s+1}^{(k)}(\alpha_{ij})}{\partial\alpha_{ij}^{l-2}}\Bigg)\;,\quad l\ge2\;.
\end{aligned}
\end{equation}
So, the terms
\[
\frac{\partial}{\partial\Sigma_n}\left(\frac{\partial^lb_s^{(k)}(\alpha_{ij})}{\partial\alpha_{ij}^l}\right)\;,\quad n=1,\ldots,5\;,
\]
can be straightforwardly computed via the chain rule when expressing the Hamilton equations in the canonical variables \eqref{eqn:Poinc} and \eqref{eqn:rescoord}.
\end{document}